# Hidden Information and Regularities of Information Dynamics II(R)

*Vladimir S. Lerner*
13603 Marina Pointe Drive, Marina Del Rey, CA 90292, USA, lernervs@gmail.com

*Abstract*
Part 1 (Lerner 2012) has studied the *conversion* of observed random process with its hidden information to related dynamic process applying entropy functional (EF), as entropy measure of the random process, and information path functional (IPF), as information measure of the dynamic conversion process. The variation principle (VP), satisfying the EF-IPF equivalence along a shortest path-trajectory, leads to information dual complementary *maxmin-minimax law*, which creates a mechanism of arising information regularities from a stochastic process.
This Part 2 studies a *mechanism* of cooperation of the observed multiple hidden information process, which follows from the law and produces cooperative structures, concurrently assembling in a hierarchical information network (IN) and generating the IN's digital genetic code. We analyze the interactive information contributions, information quality, inner time scale, information geometry of the cooperative structures; evaluate a curvature of these geometrical forms, and their cooperative information complexities. The law information mechanisms operate an information observer.
The observer, acting according the law, selects random information, converts it in information dynamics, builds the IN cooperatives, which generate the genetic code.

**Part2. Cooperative Information Dynamics and its Optimal Information Network**

*Introduction*
An observed multi-dimensional Markov diffusion process includes the impulse control, which extracts hidden information from each the process' dimensions, while transforming the Markov process to Brownian diffusion and then back to the Markov process and cuts off the process during the transformation. The cutoff selects the hidden information, concentrating in the formed Feller's kernel (Feller 1957), whose a minimal Markov path is measured by the entropy integral functional EF (Lerner 2012a). This hidden information absorbs the inner connections between the process cutoff states through all multi-dimensional process, which are arranged by a sequence of the applied impulse controls.
Equations of the minimax variation principle (VP) determine the process dynamic (macro) model with the impulse control, which converts the extracted hidden information to its equivalent dynamic information, evaluated by information path functional (IPF) defined on the VP extremal trajectories.
These equations provide a formal information mechanism, which
-selects the segments of the extremal trajectories, holding information of the cutoff segments of the random process' (micro) trajectories;
-arranges dynamic segments hierarchically according to the measured information;
-assembles these dynamic segments, carrying the ranged information, in elementary cooperative optimal structures (triplets);
- sequentially cooperates the elementary structures and integrates them in an information hierarchical network (IN), which enfolds each of its current nodes in the following node's triplet's structure with the enclosed hidden information;
-supplies the hidden information to the generated information dynamics.

For a real diffusion process with diffusing particles, the law is materialized by the particles' elementary interactions, which feed the observer with physical information substances, and realizes the IN cooperatives through real physical, chemical, biological structures that for highly organized systems would create a cognition and intelligence (Lerner 2012b). Below we describe this formal information mechanism in details with the references to Part 1(starting formulas with 1.).

## 2.1. The multi-dimensional information dynamics satisfying the VP.

Let us have a $n$-dimensional controllable dynamic process with complex conjugated eigenfunctions $\lambda_i(t) = \alpha_i(t) \pm \beta_i(t), i = 1,...,n$ of matrix $A(\lambda_i(t))$ spectrum. The stepwise controls $v = v(t_o)$, applied simultaneously at the moment $t = t_o$ to all dimensions, start dynamic process with assumed non-equal eigenvalues $\lambda_{io} = \alpha_{io} \pm \beta_{io}, i = 1,...,n$ of matrix $A_o(\lambda_{io}), i = 1,...,n$ at this moment, corresponding a pair of conjugated processes at each $k$-segment of time interval $t_k^i = \tau_k^i - \tau_{k-1}^i$, $t_o = (\tau_k^i), i = 1,...,n, k = 1,...,m$. During the dynamic movement at each segment, satisfying the VP minimax principle, each initial complex eigenvalue $\{\lambda_{io}\}, i = 1,...,n$ is transformed in real eigenvalue $\{\alpha_{it}\}, i = 1,...,n$ by the end of the segment's time interval $t_k^i$, according to the invariant relation (1.8.5):

$$\alpha_{it} = \alpha_{io} \exp \mathbf{a}_{io}(2 - \exp \mathbf{a}_{io})^{-1} , \qquad (2.1)$$

where $\mathbf{a}_{io} = \alpha_{io} t_k$ - the model's invariant, and $\{\alpha_{io}\}, i = 1,...,n$ - real components of $\{\lambda_{io}\}$, carrying the corresponding frequency of the spectrum.

Applying these optimal dynamics to macromodel (Part1), we will prove the following Proposition.

<u>Proposition 2.1.</u> Current time course of the controllable information process, satisfying the VP, is accompanied with a sequential ordering of both macromodel's information spectrum and time intervals of the extremal's segments.

Specifically, for the ranged spectrum of the model's real parts of eigenvalues $A_o(\alpha_{1o}, \alpha_{2o}, \alpha_{3o},..., \alpha_{io},..., \alpha_{no})$, selected by the VP at each process' segments, are such that $\alpha_{1o}$ holds maximal information frequency and $\alpha_{no}$ holds minimal information frequency, and it is *required to prove*:

(1) That, at given invariant $\mathbf{a}_{io}$, the $\alpha_{1o}$ is selected from the process' *shortest* segment's time interval, and $\alpha_{no}$ is selected from the process' *longest* segment's time interval; and

(2) The optimal process' current time course consists of a sequence of the segment's ordered time intervals $t_k^1, t_k^2, t_k^3,..., t_k^i,..., t_k^n$, with $t_k^1$ as a shortest segment's time interval and $t_k^n$ as a longest segment's time interval, while $t_k^1$ is the *first* time interval at the process beginning.

*Proof* (1) follows from invariant $\mathbf{a}_{io} = \alpha_{io} t_k^i$, $t_k^i = \tau_k^i - \tau_{k-1}^i$, which implies that each maximal $\alpha_{io}(\tau_{k-1}^i)$ (with a fixed $\tau_{k-1}^i$) corresponds to minimal $t_k^1$, or vice versa, each minimal $\alpha_{no}(\tau_k^n)$ is selected from maximal $t_k^n$.

*Proof* (2) is a result of reaching each local minima of the IPF functional's derivation at the moment $\tau_k^i$ of each segment's end (following from the minimum Hamiltonian, according Prop.1. 5.1):

$$E |\frac{\partial S_{ik}}{\partial t}(\tau_k^i)| = |\alpha_{ik}(\tau_k^i)| \to \min, i = 1,...,n, \qquad (2.2)$$

(along time course of the optimal process with $(\tau_k^i, \tau_{k+1}^{i+1},...\tau_m^n), i=1,...n, k=1,...,m$)) and reaching a global minima for this functional's derivations at the process end:

$$E |\frac{\partial S_{kn}}{\partial t}| = |\sum_{i=1}^{n} \alpha_{ik}| \to \min. \qquad (2.2a)$$

Holding minimal derivation (2.2), invariant preserves this minimum along each segment for each fixed $\alpha_{io}$. Minimal path functional on each dimension's fixed time interval $t_k^i = \tau_k^i - \tau_{k-1}^i$ brings minimal increments of this functional and adds it to each following segment's minima according to (2.2), allowing to reach (2.2a) by the end.

Each current minimal eigenvalue, selected by the VP minimax principle from this multi-dimensional spectrum, has a maximum among other minimal eigenvalues.

From that it follows a sequential declining of these minimal increments at moments $(\tau_k^i, \tau_{k+1}^{i+1},...\tau_m^n), i=1,...n, k=1,...,m$ along the time course of the optimal process:

$$\min \alpha_{i+1t}(\tau_{k+1}^{i+1}) < \min \alpha_{it}(\tau_k^i) \text{ or } \min \alpha_{it}(\tau_k^i) > \min \alpha_{i+1t}(\tau_{k+1}^{i+1}), \tau_{k+1}^{i+1} > \tau_k^i,.... \qquad (2.3)$$

In such sequence, each minimal eigenvalue $\alpha_{it}(\tau_k^i)$ holds a maximum regarding all following minimal eigenvalues.

From (2.1, 2.2), it follows that minimum of $\alpha_{it}$ leads to minimum for $\alpha_{io}$ (at a fixed invariant).

Since each sequence of $\alpha_{it}(\tau_k^i)$ corresponds to related sequence of minimals $\alpha_{io}(\tau_{k-1}^i)$ along the time course of the optimal process $\alpha_{io}(\tau_k^i, \tau_{k+1}^{i+1},...\tau_m^n)$, both $\alpha_{it}(\tau_k^i)$ and $\alpha_{io}(\tau_{k-1}^i)$ will be orderly arranged during this time course. Then, this ordered connection holds true for all $A_t(\alpha_{1k}, \alpha_{2k}, \alpha_{3k},...,\alpha_{ik},...,\alpha_{nk})$ and related $A_o(\alpha_{1o}, \alpha_{2o}, \alpha_{3o},...,\alpha_{io},...,\alpha_{no})$ during the optimal process' time course. Sequential ordering of both eigenvalues $\alpha_{it}(\tau_k^i)$ and $\alpha_{io}(\tau_{k-1}^i)$ leads to ordering of the corresponding segment's time intervals $\{t_k^i\}$, starting with its minimal $t_k^1$ at the process beginning with its maximal $\alpha_{1o}$.

The proposition is proved, confirming also the initial assumption of the ranged spectrum. •

*Therefore,* ordering of the initial eigenvalues satisfies to the minimax principle, which allows selecting sequentially such following eigenvalue for this optimal spectrum that brings a maximal eigenvalue among all other minimal eigenvalues of the spectrum.

## 2.2. Forming an optimal cooperative information dynamic structure

The minimax principle imposes the constraint and control that connect the extremal segments, leading to the process' cooperative dynamics, which includes both physical and virtual forms.
Here we focus on the *optimal* conditions of segment's cooperation, following from the minimax.
According to Prop.2.1 and (2.1), the sequential arrangement of the segments accompanies with decrease of their starting and ending eigenvalues. Because of simultaneous start of all spectrums' initial eigenfunctions, the segments' might cooperate only through joining their *ending* eigenvalues, which could initiates the cooperative motion. Such possibility arises by the end of each segment's local dynamics with starting the between segments information dynamics.

Since the time intervals of the local dynamics are different, a minimal time of a segment's cooperation with other consecutive segment cannot be less than time internal of completion this segment's optimal dynamics. In addition, any elementary cooperation of two segments requires equalization of the segments' ending eigenvalues before joining them together.

Theoretically, reaching the equalization could be possible during a time interval between completions of these segments' dynamics: $\Delta t_k^i = t_k^{i+1} - t_k^i$, where $t_k^i, t_k^{i+1}$ are time intervals for these segments' ending eigevalues $\alpha_{it}(t_k^i)$ and $\alpha_{i+1t}(t_k^{i+1})$ at $\alpha_{it}(t_k^i) > \alpha_{i+1t}(t_k^{i+1})$.

In such an optimal arranged spectrum, each its previous segment would require minimal time interval $\Delta t_k^i$ needed to be spent on the equalization with the following segment. For example, three sequentially cooperating segments with decreasing ending eigenvalues $\alpha_{it}(t_k^i), \alpha_{i+1t}(t_k^{i+1}), \alpha_{i+2t}(t_k^{i+2})$ need two of such time intervals $\Delta t_k^i = t_k^{i+1} - t_k^i$ and $\Delta t_k^{i+1} = t_k^{i+2} - t_k^{i+1}$, where $t_k^{i+2}$ is the time interval, which is necessary to finish optimal dynamics on third segment.

A total minimal time interval for equalization of two eigenvalues with that in the third segment: $\Delta t_k^i + \Delta t_k^{i+1} = t_k^{i+2} - t_k^{i+1} + t_k^{i+1} - t_k^i = t_k^{i+2} - t_k^i$, being added to the first $t_k^i$: $\Delta t_k^{i+1} + t_k^i = t_k^{i+2} - t_k^i + t_k^i = t_k^{i+2}$, should be equal to the time interval, required by the third segment's dynamics $t_k^{i+2}$.

These two steps' consolidation includes joining first the pair of segments, having decreasing maximums of their eigenvalues, and then assembling the joint pair to the third segment with a minimal eigenvalue (Fig.2.1).

Performing this cooperation requires changing the sign of starting stepwise control on the third segment to be opposite to the signs of both first and the second segments' starting controls.

Then, to apply the impulse control to each of these two segments (at the moments $t_k^{i1}, t_k^{i+1}$ accordingly) for changing the signs of their eigenfunctions, providing their decrease toward equalization of the eigenvalues by moment $\tau_k^{i+1}$. At such first cooperation, the joint minimal eigenvalue equals to $2\alpha_{i+1t}(\tau_k^{i+1})$ will be reached, and then the joint eigenvalue $3\alpha_{i+2t}(\tau_k^{i+2})$ after the second cooperation is achieved. If the first cooperation would coincide with the second one by cooperating all three eigenvalues simultaneously at the moment $\tau_k^{i+2}$ with forming $3\alpha_{i+2t}(\tau_k^{i+2})$, then the second cooperation with $2\alpha_{i+1t}(\tau_k^{i+1})$ is not needed.

It means, a sum of the derivations, according to (2.2a), would be minimal: $E|\frac{\partial S_{i,i+1,i+2}}{\partial t}| = 3|\alpha_{i+2t}(\tau_k^{i+2})| < |2\alpha_{i+1t}(\tau_k^{i+1}) + 3\alpha_{i+2t}(\tau_k^{i+2})|$ .

In this case, the ending eigenvalue in the second cooperation will get maximal decrease $\alpha_{it}(\tau_k^i) - \alpha_{i+2t}(\tau_k^{i+2})$ during time interval $\Delta t_k^{i+2}$, compared to its decrease $\alpha_{it}(\tau_k^i) - \alpha_{i+1t}(\tau_k^{i+1})$ during time interval $\Delta t_k^i$ for the first cooperation. At $\alpha_{i+2t}(\tau_k^{i+2}) < \alpha_{i+1t}(\tau_k^{i+1})$, the sum $2\alpha_{i+1t}(\tau_k^{i+2})$ taken by the moment $\tau_k^{i+2}$ would be minimal comparing it with that at the moment $\tau_k^{i+1}$: $2\alpha_{i+1t}(\tau_k^{i+2}) < 2\alpha_{i+1t}(\tau_k^{i+1})$. This means, shifting a pair cooperation from moment $\tau_k^{i+1}$ to the moment $\tau_k^{i+2}$ will bring less derivations (2.2a) just for this pair. It follows that *only the triple cooperation* at the moment $\tau_k^{i+2}$ satisfies to the minimax principle. Moreover, since optimal cooperation completes during the minimal time interval of the optimal dynamics on the third segment, this segment does not require additional time for equalization and cooperation.

Implementation of this procedure requires determining such an ordered location of each three triplet's segments (within the system's arranged spectrum) that allow finishing cooperation of two triplet's segments with a third triplet's segment by the end of the dynamic process on third segment.

<u>Comments 2.1.</u> Joining the equal eigenvalues for a pair of segments in both sides of equality

$$\alpha_{it}(\tau_k^i) - \alpha_{i+2t}(\tau_k^{i+2}) = \alpha_{it}(\tau_k^i) - \alpha_{i+1t}(\tau_k^{i+1}) + \alpha_{i+1t}(\tau_k^{i+1}) - \alpha_{i+2t}(\tau_k^{i+2}) \quad (2.4)$$

at $\Delta t_k^{i+1} = t_k^{i+2} - t_k^i = \Delta t_k^i + \Delta t_k^{i+1}$ (2.4a)

with their equivalent segments' information $\alpha_{it}(\tau_k^i)\Delta t_k^{i+2} = \alpha_{it}(\tau_k^i)(\Delta t_k^i + \Delta t_k^{i+1}))$, (2.4b) requires spending the control's quantity of information $\mathbf{a}_o^2(\gamma)$ on the segment's *cooperation*.

This quantity information is supplied by a *hidden* information between information states that had been enclosed in the entropy functional before the cutoff (Sec.1.3). •

Because all segments of the optimal ensemble are similar, the optimal cooperation of their ranged eigenvalues, satisfying the VP, requires cooperating the optimal ensemble by the triplet's units, being arranged sequentially. To continue the cooperation from the very first triplet, its ending segment (with the enclosed triplet's information) should be joint to the two following segments (in their ranged sequence), forming a next triplet, whose ending segment will be joint to the following next two segments of this sequence, and so on (Fig.2.1). Since both optimal ranging and optimal cooperation of the eigenvalues should satisfy the minimax principle, they could proceed concurrently during the time course of optimal process, implementing consecutively this principle.

This means that after starting the model's optimal process in its time course, a first maximal eigenvalue should be chosen from a minimal time interval of the time course. Then, following the time course, a next chosen nearest minimal time intervals determines the next maximal eigenvalue in the sought ranged sequence. While searching for the cooperation with a third segment (in this sequence), it is assumed that dynamic processes in both previous segments had already started. And their consolidation dynamics, directed on consolidation with the third one, supposed to be finished by the moment of finding third segment. Each selection, satisfying the minimax principle, is applying sequentially to a next stage of both ranging eigenvalues and their ordered cooperation.

In such optimal cooperative dynamics, cooperation of all triplets' units of the model's segments completes within the time interval of the last segment's dynamics that belongs to an ending triplet.

Let us consider a balance of information for a triplet, assuming that each segment time interval retains $\mathbf{a}_o(\gamma)$ units of information, stepwise control brings information $\mathbf{a}(\gamma)$, and impulse control delivers the information contribution $\mathbf{a}_o^2(\gamma) \cong 2\mathbf{a}(\gamma)$. To start a first three segment, three stepwise controls are needed that brings $3\mathbf{a}(\gamma)$ information units. These segment's dynamics retain $3\mathbf{a}_o(\gamma)$ information units. To initiate the fist and the second segment's dynamics, directed on their consolidation with the third segment, two impulse controls with information $2\mathbf{a}_o^2(\gamma)$ should be applied for connecting the segment's ending eigenvalues $\alpha_{it}(\tau_k^i), \alpha_{i+1t}(\tau_k^{i+1})$ with the eigenvalues $\alpha_{it}(\tau_{k+1}^i), \alpha_{i+2t}(\tau_{k+1}^{i+2})$ on the next two time intervals $\Delta t_k^{i,3}, \Delta t_k^{i+1,3}$.

The related quantities of information are $\alpha_{it}(\tau_{k+1}^i)\Delta t_k^{i,3} + \alpha_{i+2t}(\tau_{k+1}^{i+2})\Delta t_k^{i+1,3}$.

After reaching equalization with the third segment's eigenvalue, two of the impulse controls should be applied to make the segment's joint connection (one control-for joining two first segments in a duplet, another one- to join this doublet with the third segment). Both impulse controls contribute additional $2\mathbf{a}_o^2(\gamma)$ units of information, while the cooperation consumes information $-(\alpha_{it}(\tau_{k+1}^i)\Delta t_k^{i,3} + \alpha_{i+2t}(\tau_{k+1}^{i+2})\Delta t_k^{i+1,3}) \cong 2\mathbf{a}(\gamma) \cong \mathbf{a}_o^2(\gamma)$ which is equivalent to the single unit the impulse' control information. We come to balance Eq in the form

$$3\mathbf{a}(\gamma) - 3\mathbf{a}_o(\gamma) + 2\mathbf{a}_o^2(\gamma) - (\alpha_{it}(\tau_{k+1}^i)\Delta t_k^{i,3} + \alpha_{i+2t}(\tau_{k+1}^{i+2})\Delta t_k^{i+1,3}) + 2\mathbf{a}_o^2(\gamma) \cong 2\mathbf{a}(\gamma). \qquad (2.5)$$

Satisfaction of this equality is confirmed in (Lerner 2008a) by a direct computation.

Since third segment of the fist triplet serves as a first segment of a next triplet to be sequentially formed, it will cooperate with the following two segments of the model's spectrum, carrying its surplus information $\mathbf{a}_o^2(\gamma)$ (at $\mathbf{a}(\gamma) + \mathbf{a}_o(\gamma) \cong 2\mathbf{a}_o^2(\gamma)$) to the next triplet formation.

We get the balance Eq for the second triplet in the form

$$2\mathbf{a}(\gamma) + 2\mathbf{a}_o^2(\gamma) + \mathbf{a}_o^2(\gamma) - 2\mathbf{a}_o(\gamma) - 2\mathbf{a}(\gamma) \cong 0. \qquad (2.5a)$$

Hence, the second triplet does not need additional external information for the cooperation, its cooperative two units of impulse control $2\mathbf{a}_o^2(\gamma)$ includes one unit $\mathbf{a}_o^2(\gamma)$, delivered from the fist triplet, another unit carries the second segment of this forming triplet. However, formation of the third triplet requires the same two units of the impulse control $2\mathbf{a}_o^2(\gamma)$ as that for forming the first triplet. This leads to balance Eq. (2.5), which transfers its surplus $\mathbf{a}_o^2(\gamma)$ to the next forth triplet, and so on. Thus, each three segments, acting separately, need external information $3\mathbf{a}(\gamma) + 3\mathbf{a}_o^2(\gamma)$, whereas their cooperation requires additional information $\mathbf{a}_o^2(\gamma)$.

The balance Eq. for each separated segment: $\mathbf{a}_{oi}^2(\gamma) + \mathbf{a}_i(\gamma) - \mathbf{a}_{oi}(\gamma) = \delta \mathbf{a}_i(\gamma)$ is satisfied with error $\delta \mathbf{a}_i(\gamma)$, whose relative value $\delta \mathbf{a}_i(\gamma) / \mathbf{a}_{oi}(\gamma)) = \delta^*(\gamma)$ depends on $\gamma: \delta^*(\gamma \to 0) \cong 0.07, \delta^*(\gamma = 0.8) \cong 0.066$. The error $\delta \mathbf{a}_i(\gamma)$, being relative to the cooperative contribution: $\delta \mathbf{a}_i(\gamma) / \mathbf{a}_{io}^2(\gamma) \cong \delta_a^*(\gamma)$, takes a minimal surplus $\delta_a^*(\gamma \to 0) \cong 0.09$, while each segment shares only $1/3\delta_a^*(\gamma)$.

*Let us summarize* the *optimal* conditions of segment's cooperation, imposed by the minimax principle.

1. Under starting stepwise control, applied simultaneously to *n*-dimensional dynamics, the segments are orderly ranged during the optimal motion, providing the implemention of this principle;

2. The ordered assembling of the segments in the optimal triplets keeps such arranged sequence of the initial eigenvalues, which allows each two triplet's segments to finish their cooperation with a third triplet's segment by the end of the dynamic process on this segment.

This requirement determines the specific ratios of each triplet's initial eigenvalues $\gamma_1^\alpha = \alpha_{io}/\alpha_{i+1o}$, $\gamma_2^\alpha = \alpha_{i+1o}/\alpha_{i+2o}$ satisfying the invariant relations:

$$\gamma_1^\alpha = \frac{\exp(\mathbf{a}(\gamma)\gamma_2^\alpha) - 0.5\exp(\mathbf{a}(\gamma))}{\exp(\mathbf{a}(\gamma)\gamma_2^\alpha/\gamma_1^\alpha) - 0.5\exp(\mathbf{a}(\gamma))}, \gamma_2^\alpha = 1 + \frac{\gamma_1^\alpha - 1}{\gamma_1^\alpha - 2\mathbf{a}(\gamma)(\gamma_1^\alpha - 1)}, \quad (2.6)$$

where invariants $\mathbf{a}(\gamma)$ can be found from Eq (1.8.8) using a known invariant $\mathbf{a}_o(\gamma)$ which depends on $\gamma$ according to (1.8.4). Relation (2.6) follows from imposing above requirements on forming each triplet from three $(i, i+1, i+2)$ segments:

$$\alpha_{it}(\tau_k^i)\exp(\alpha_{it}(\tau_k^i)\Delta t_k^{i,3})[2 - \exp(\alpha_{it}(\tau_k^i)\Delta t_k^{i,3})]^{-1} = \alpha_{i+2t}(\tau_k^{i+2}), \text{ and}$$

$$\alpha_{i+1t}(\tau_k^{i+1})\exp(\alpha_{i+1t}(\tau_k^{i+1})\Delta t_k^{i+1,3})[2 - \exp(\alpha_{i+1t}(\tau_k^{i+1})\Delta t_k^{i+1,3})]^{-1} = \alpha_{i+2t}(\tau_k^{i+2}), \alpha_{it}\tau_k^i = \alpha_{i+1t}\tau_k^{i+1} = \mathbf{a}_i = \mathbf{a}(\gamma).$$

Condition (2.6) of the eigenvalues' cooperation in the triplet's forms limits the diapason of admissible $\gamma \to (0.0 - 0.8)$, where for the optimal dynamics with $\mathbf{a}_i(\gamma \to 0) \cong 0.23$ we get $\gamma_1^\alpha \cong 2.460, \gamma_2^\alpha \cong 1.817$.

The equalization of all three triplet's eigenvalues takes place not in a single point at the same time, but rather in a small region, forming an uncertainty zone UR (at the windows), Sec.1.11. For the triplet's ranged eigenvalues, a relative size of this zone is evaluated by an average quadratic error:

$$\varepsilon(\gamma)^2 = [(\gamma_2^\alpha(\gamma))^{-2} - (\gamma_1^\alpha(\gamma)\gamma_2^\alpha(\gamma))^{-1}], \varepsilon = (\delta t_{i+2o}/t_{i+2o}), \quad (2.6a)$$

which follows from this zone's time interval: $\delta t_{i+2o} = [(t_{i+1o})^2 - t_{io}t_{i+2o}]^{1/2}$, expressed in relative form:

$(\delta t_{i+2o}/t_{i+2o})^2 = (t_{i+1o}/t_{i+2o})^2 - (t_{io}/t_{i+2o}), t_{i+1o}/t_{i+2o} = \gamma_1^\alpha(\gamma)/\gamma_2^\alpha(\gamma), t_{i+2o}/t_{io} = \gamma_2^\alpha(\gamma)$.

For example, at average $\gamma = 0.5$, $\gamma_1^\alpha = 2.2155, \gamma_2^\alpha \cong 1.7583$, we get $\varepsilon^2 \cong 0.0675$; and at $\gamma \to 0$, with $\gamma_1^\alpha \cong 2.460, \gamma_2^\alpha \cong 1.817$ we get $\varepsilon^2 \cong 0.079$. At $\gamma \to 0$, $\gamma_1^\alpha \cong 2.460, \gamma_2^\alpha \cong 1.817$.

Since, a junction of three segments forms the UR, it corresponds engagement of two punched localities, with quantum dynamics at both windows' border (Lerner 2012). External information is delivered to the widows by two impulse controls. After the eigenvalues' equalization in the two localities, the impulse controls lock both windows with merging three segments into a triplet as a single unit. With above evaluation of widow's UR zone: $\varepsilon(\gamma)^2/2 \cong 0.033$, $\varepsilon_k \cong 0.1816$ estimates an average interval of applying impulse control, or average admissible error.

Comments 2.2

Forming a cooperative requires the existence of *cooperative information force* between the potential cooperating segments.

Let us find it using relations for a gradient of information $I_{ki}^\alpha$ between the segments $k, i$:

$$X_{ki}^{\alpha} = -\frac{\delta I_{ki}^{\alpha}}{\delta l_{ki}^{\alpha}}, \tag{2.7}$$

where $\delta l_{ki}^{\alpha}$ is a difference in positions of the segments $k,i$ at $\partial l_{ki}^{\alpha} = \delta t_{ki}^{\alpha} c_{ok}$, $\delta t_{ki}^{\alpha}$ is the related time shift, and $c_{ok}$ is segments' space speed. We apply (2.7) considering an increment of information between two segments $k,i$ at some current moment $t_k : \delta I_{ki}(t_k)$ at a finite intervals between segments $\partial l_{ki}^{\alpha} = l_k^{\alpha} - l_i^{\alpha} = \delta t_{ki}^{\alpha} c_{ok} = (t_k^{\alpha} - t_i^{\alpha}) c_{ok}$ and a fixed speed, where $\delta I_{ki}(t_k)$ depends on the difference of contribution from both segments $k : \alpha_{\tau k}(t_k^{\alpha}) t_k^{\alpha}$ and contribution from segments $i$, collected by moments $t_k$ during time interval $t_{ki} = t_k^{\alpha} - t_i^{\alpha}$ between these segments: $\alpha_{\tau i}(t_i^{\alpha})(t_k^{\alpha} - t_i^{\alpha})$. We get

$$\delta I_{ki}(t_k) = \alpha_{\tau k}(t_k^{\alpha}) t_k^{\alpha} - \alpha_{\tau i}(t_i^{\alpha})(t_k^{\alpha} - t_i^{\alpha}). \tag{2.7a}$$

The increment between segment $l,i$ at $t_{li} = t_l^{\alpha} - t_i^{\alpha}$ by analogy is

$$\delta I_{li}(t_{li}) = \alpha_{\tau l}(t_l^{\alpha}) t_l^{\alpha} - \alpha_{\tau i}(t_i^{\alpha})(t_l^{\alpha} - t_i^{\alpha}).$$

Both increment determine the finite cooperative forces between these segments

$$\delta X_{ki}^{\alpha} = X_{ki}^{\alpha} \delta l_{ki}^{\alpha} = -\delta I_{ki}(t_k), \delta X_{li}^{\alpha} = X_{li}^{\alpha} \delta l_{li}^{\alpha} = -\delta I_{li}(t_{ki}).$$

The relative increments to the first segments:

$$\delta I_{ki}(t_{ki})/t_i^{\alpha} = \alpha_{\tau k}(t_k^{\alpha}) t_k^{\alpha}/t_i^{\alpha} - \alpha_{\tau i}(t_i^{\alpha})(t_k^{\alpha}/t_i^{\alpha} - 1), \delta I_{li}(t_{li})/t_i^{\alpha} = \alpha_{\tau l}(t_l^{\alpha}) t_l^{\alpha}/t_i^{\alpha} - \alpha_{\tau i}(t_i^{\alpha})(t_l^{\alpha}/t_i^{\alpha} - 1)$$

and their ratio acquires the form

$$\delta I_{kli}^{*} = \delta I_{ki}(t_{ki})/\delta I_{li}(t_{li}) = \frac{\alpha_{\tau k}(t_k^{\alpha}) \gamma_{ki}^{\alpha} - \alpha_{\tau i}(t_i^{\alpha})(\gamma_{ki}^{\alpha} - 1)}{\alpha_{\tau l}(t_l^{\alpha}) \gamma_{li}^{\alpha} - \alpha_{\tau i}(t_i^{\alpha})(\gamma_{li}^{\alpha} - 1)} = \frac{2 - \gamma_{ki}^{\alpha}}{2 - \gamma_{li}^{\alpha}}. \tag{2.7b}$$

We get the ratios for a triplet with different $\gamma_{li}^{\alpha}(\gamma)$:

$\gamma_{ki}^{\alpha}(\gamma \to 0) = 2.46, \gamma_{li}^{\alpha}(\gamma \to 0) = 4.794, \delta I_{kli}^{*} = 0.1645;\quad \gamma_{ki}^{\alpha}(\gamma = 0.5) = 2.215, \gamma_{li}^{\alpha}(\gamma = 0.5) = 3.895, \delta I_{kli}^{*} = 0.113;$

$\gamma_{ki}^{\alpha}(\gamma = 0.8) = 1.969, \gamma_{li}^{\alpha}(\gamma = 0.8) = 3.3, \delta I_{kli}^{*} = -0.0235.$

These relative increments decrease with growing $\gamma$, while the last triplet fails to cooperate. The increments evaluate information cooperative forces at time interval starting the triple cooperation. By the end of cooperation, each increment at $t_{ki} \to 0, t_{li} \to 0$, acquires invariant measure $\alpha_{\tau k}(t_k^{\alpha}) t_k^{\alpha} = \alpha_{\tau l}(t_l^{\alpha}) t_l^{\alpha} = \mathrm{a}_{\tau}$ and their ratio $\delta I_{kli}^{*} \to 1$. It's seen that cooperative increment between first pair of segments is less than that for the second pair, depending of the limited segment's ratio determined by admissible $\gamma$.

Information potential of current IN triplet $m_i$, currying information $\delta I_{ik}^m = \mathbf{a}_i + \mathbf{a}_{oi}^2 \cong \mathbf{a}_{oi}$, to attract $m_k$ triplet depends on both $\mathbf{a}_{oi}(\gamma)$ and on relative distance $(t_i^m - t_k^m)/t_i^m = (l_i^m - l_k^m)/l_i^m = \partial l_{ik}^{m*}$:

$$X_{ik}^{Im} = -\frac{\delta I_{ik}^m}{\delta l_{ik}^{\alpha}} = \mathbf{a}_{oi}(\gamma)(\gamma_{ik}^m - 1), \tag{2.8}$$

which measures this potential directly in Nats (bits). The information potential, relative to information of the first triplet, determines cooperative force between these triplets:

$$X_{1k}^{Im1} = (\gamma_{1k}^m - 1) \tag{2.8a}$$

Comments 2.3. Since each pair cooperation requires $\mathbf{a}_{oi}^2(\gamma) \cong 2\mathbf{a}_i(\gamma)$ unit of information and time of cooperation $t_{ic}$, we may determine a *speed of cooperation*

$$c_{ic} = 2\mathbf{a}_i(\gamma)/t_{ic}, \qquad (2.9)$$

which, depending on $t_k^{i+2} \geq t_{ic} \geq \Delta t_k^{i+2}$, can take minimal or maximal values, while the minimal

$$c_{ico} = 2\mathbf{a}_i(\gamma)/t_k^{i+2} = 2\mathbf{a}_i(\gamma)/\mathbf{a}_{io}(\gamma)\alpha_{i+2o}(t_k^{i+2}) \qquad (2.9a)$$

limits cooperation through the admissible information speed on third segment $\alpha_{i+2o}(t_k^{i+2})$. For $t_{ic} = \Delta t_k^{i+2} = t_k^{i+2} - t_k^i = (\gamma_i^\alpha - 1)t_k^i$ we get maximal speed

$$c_{ico} = 2\mathbf{a}_i(\gamma)/\mathbf{a}_{io}(\gamma)(\gamma_i^\alpha - 1)\alpha_{io}(t_k^i), \qquad (2.9b)$$

evaluating the cooperative speed via the initial eigenvalue for each triple.

At $\Delta t_k^{i+2}/t_k^{i+2} = \varepsilon_k$, or if for some $m$-th triplet its $\Delta t_k^{i+2} = \Delta t_k^m$ reaches minimal limited $\delta t_k^{\min}$, admissible by average error $\varepsilon_m = \delta t_m^{\min}/t_m$, then the ratio of the maximal to minimal cooperative speed for a triplet is $C_c^* = c_{ico}/c_{ic} = (\varepsilon_k)^{-1}$, or at $\varepsilon_k \cong 0.1816$ this ratio is limited by $C_c^* \leq 5.5$.

This speed restricts formation of the IN information logic (Lerner2010a). The IN ending eigenvalue is determined through initial eigenvalue of the first IN triplet at $i = m : \alpha_{mo}(t_k^m) = \alpha_{1o}(t_k^1)/(\gamma_1^\alpha)^m$.

Ratio $\alpha_{i=1,o}^{m=1}/c_{ico} = C_{oc}$ is a cooperative analog of the ratio of the IN maximal signal speed to the current cooperative-encoding speed, which evaluates the IN relative channel capacity: $C_{oc} \cong 1/2\mathbf{a}_{io}(\gamma)\mathbf{a}_i^{-1}(\gamma)(\gamma_{i=m}^\alpha - 1)(\gamma_{i=m}^\alpha)^m$.

At $\gamma = 0.5, \gamma_i^\alpha \cong 4$, we estimate it by $C_{oc} \cong 5.23 \times 4^m$, which depends on maximal number of the IN enclosed nodes on the hierarchical levels $m = n/2$, or the ability to cooperate these numbers for observing systems' dimension $n$. The observer's cooperative bandwidth limits both number $n$ and $m$. For example at $m = 8$, $C_{oc} \cong 2.25 \times 10^{10}$, to cooperate $c_{ico} = 1 bit/\sec$ requires starting frequency $B_f \cong 2.25 \times 10^{10} bit/\sec \approx 22.5 Gbit/\sec$. This needs simultaneously applied $m = 8$ impulse controls, whose extracting information will be cooperating in the IN by the same impulse controls through their reflected inner controls. According to Koch et al, 2006, a human eye's retina holds $\alpha_{i=1,o}^{m=1} \approx 10^6 bit/s$, while a single neuron low speed is $c_{hco} \approx 10 bit/s$, which limit the cooperative speed. We get $C_{oc} \cong 10^5$, and $m = 7.3 \cong 7, n = 14$. Hence, a complete IN of human being information logic can be build by its sections, each with $m \cong 7$ levels, which cooperate in each triplet of a future IN, composing all observed integral information (Fig.2.4). At $c_{hco} \approx 10 bit/s$, these lead to estimation

$$B_f \cong 10^6 bit/s \approx 10^{-3} Gbit/s. \bullet \qquad (2.10)$$

Comments 2.4

The triple cooperation keeps optimality and the stability for each current triple as well as for each of their ranged sequence. Adding to the triple a forth-ordered eigenvalue will make the cooperation of this quadruple impossible, because, at such interaction, the cooperation of more than three elements becomes *unstable* (Lichtenberg, Lieberman 1983).

For any presumable ordered optimal manifold of eigenvalues with *m simultaneous and potentially stable co operations*, the minimal dynamic path (MP) (from every cooperation) up to a final cooperation would not exceed that with the time interval, limited by a minimal final eigenvalue $\alpha_{mo}$ in this manifold. The cooperation, performed *sequentially* by triple or double, will hold the same minimal MP, preserving the stability at each of such elementary cooperation.

For example, at *m* =4, the two sequential double co-operations require the minimal time interval limited by $\alpha_{4o}$, which is the same if the simultaneous cooperation of these four segments would be stable. For *m* =5, the two triple sequential co-operations (Fig.2.1) during the time interval, limited by minimal $\alpha_{5o}$, is the same if the simultaneous cooperation of these five segments would be stable.

Theoretically, cooperation of four segments in a quadruple simultaneously is possible: by making sequentially three pair connections, or forming different combinations of triple connections. If each of the quadruples will satisfy to minimax, then the second of such joint quadruples would have a minimal time interval of its seventh'(final segment) dynamics, which is equal to a total minimal time interval of both quadruples jointly. Three optimal simultaneously joint triplets with the seventh final segment have the same minimal time interval.

In this consideration, both of these information structures: three triplets and two quadruples sound equivalent, both satisfy minimax and have the same total minimal time interval.

Let us analyze each of them. Number of combinations to make pair connections for a triplet is *three*. Therefore, three triplets would require *nine* such combinations. Number of combinations to make pair connections for a quadruple is *six*. Hence, two of these quadruples require 12 pair connections. Therefore, a choice of these connections has uncertainty for the two quadruples in 12/9 $\approx$ 1.33 times higher than that for the tree triplets. For the considered possibility of triple connections: a quadruple requires *four* number of combination for triple connections, and a triplet needs only *one* such combination. Hence, the choice of these connections has a relative uncertainty 8/3 $\approx$ 2.66, which is in two times higher than that for their comparative pair connections.

The results show that forming optimal triplet requires minimum uncertainty (entropy) than any of quadruples, even without reference to stability of each connection.

In addition to that, since each pair connection requires impulse control information $\mathbf{a}_o^2(\gamma)$, the quadruple would spend $6\mathbf{a}_o^2(\gamma)$ compare to the triplet's spending $2\mathbf{a}_o^2(\gamma)$. •

Therefore, the minimal MP is achieved for an elementary cooperative unit with a *maximum* of *three* eigenvalues (with a minimal time), where for a space distributed macromodel, a *minimal* number of cooperating segments is three, each one for every space dimension.

## 2.3. *The interactive information contribution of the process' inner controls, asymmetry, information quality and inner time scale.*

Here we deal with an observer inner control, as interactive reaction of external impulse, applied according to the minimax principle.

Considering dynamic form of information functional (Part1):

$$S[x(t)] = \int_s^T 1/2(a^u)^T (2b)^{-1} a^u dt,$$

with related dynamic equations

$$a_i^u = \lambda_i(x_i + v_i), (a_i^u)^T = \lambda_i^*(x_i^* + v_i^*), 2b_i = \dot{r}_i, r_i = x_i x_i^*, 2b_i = (\dot{x}_i x_i^* + \dot{x}_i^* x_i) = \lambda_i x_i x_i^* + \lambda_i^* x_i^* x_i$$

and information Hamiltonian

$$H_i^S = \frac{\lambda_i(x_i + v_i)(x_i^* + v_i^*)\lambda_i^*}{2(\lambda_i + \lambda_i^*)x_i x_i^*} \quad \text{at } x_i x_i^* = x_i^* x_i. \tag{3.1}$$

By eliminating in $H_i^S$ the direct impact of conjugated and non-conjugated controls on the state, which they do not control: $v_i^* x_i + v_i x_i^* = 0$, we get Hamiltonian, measuring the mutual interactive information delivered by control:

$$H_i^{S_u} = \frac{\lambda_i \lambda_i^* v_i v_i^*}{2(\lambda_i + \lambda_i^*)x_i x_i^*}. \tag{3.1a}$$

Let us apply optimal control $v_i^o = -2x_{io}(t_{io}), v_i^{o*} = -2x_{io}^*(t_{io}^*)$ during time interval $t_i$ and $t_i^*$ accordingly.

At the initial moment $t_{io} = t_{io}^*$ of applying this control, the Hamiltonian holds

$$H_i^{S_u} = \frac{\lambda_{io} \lambda_{io}^*}{2(\lambda_{io} + \lambda_{io}^*)}, \text{ where at } \lambda_{io} = \alpha_{io} + j\beta_{io}, \lambda_{io}^* = \alpha_{io} - j\beta_{io}, \frac{v_i^o v_i^{o*}}{x_{io} x_{io}^*} = 4, \text{ we get}$$

$$H_i^{S_u} = H_i^{S_u} = \frac{(\alpha_{io}^2 + \beta_{io}^2)}{\alpha_{io}}, \text{ or } H_{io}^{S_u} = \alpha_{io}(1+\gamma_{io}^2) \text{ at } \gamma_{io} = \beta_{io}/\alpha_{io}. \tag{3.2}$$

Let us determine the Hamiltonian at the ending moment of the applied optimal control.

We get the related increments for $x_{it} = x_{io}[2-\exp(\lambda_{io}t_i)]$, $x_{it}^* = x_{io}^*[2-\exp(\lambda_{io}^* t_i^*)]$ and $\lambda_{it} = \lambda_{io}[2-\exp(\lambda_{io}t_i)]^{-1}, \lambda_{it}^* = \lambda_{io}^*[2-\exp(\lambda_{io}^* t_i^*)]^{-1}$.

Considering the Hamiltonian at fixed initial $\frac{v_i^o v_i^{o*}}{x_{io} x_{io}^*} = 4$ we have

$$H_{io}^{S_u} = \alpha_{it}(1+\gamma_{it}^2), \gamma_{it} = \beta_{it}/\alpha_{it}, \text{ or } H_{it}^{S_u} = \alpha_{io}[2-\exp(\alpha_{io}t_i)]^{-1}(1+\gamma_{it}^2). \tag{3.3}$$

This increment during interval $t_i$ holds the entropy contribution in a form of step-function: $\Delta S_{io}^{S_u} = \alpha_{io} t_i [2-\exp(\alpha_{io}t_i)]^{-1}(1+\gamma_{it}^2)$, or using dynamic invariant $a_o = \alpha_{io} t_i$ we get $\Delta S_{it}^{S_u} = a_o [2-\exp(a_o)]^{-1}(1+\gamma_{it}^2)$.

Coefficient $\gamma_{it}$ decreases during the optimal process from a fixed $\gamma_{io}$ to $\gamma_{it} \to 0$ since $\beta_{it} \to 0$ by the end of $t_i$. That is why the entropy contribution holds $\Delta S_{it}^{S_u} \to a_o [2-\exp(a_o)]^{-1}$, or at optimal $a_o(\gamma_{io} \cong 0.5) \approx 0.75$ we get $\Delta S_{it}^{S_u} \approx 0.5$, which is estimated by $\Delta S_{it}^{S_u} \approx a_o^2$. This relation evaluates interaction of the controls, applied to an extremal segments with only real eigenvalues. While the interactive control's Hamiltonian (3.1) counts also imaginary component of the conjugated process.

Let us find Hamiltonian for more general case, at any current moment $t_i$ within interval of applying control:

$$H_i^{S_u} = \frac{\lambda_{io}[2-\exp(\lambda_{io}t_i)]^{-1}\lambda_{io}^*[2-\exp(\lambda_{io}^*t_i)]^{-1}v_i(t_i)v_i^*(t_i)}{2(\lambda_{io}[2-\exp(\lambda_{io}t_i)]^{-1}+\lambda_{io}^*[2-\exp(\lambda_{io}^*t_i)]^{-1})x_{io}[2-\exp(\lambda_{io}t_i)]x_{io}^*[2-\exp(\lambda_{io}^*t_i)]}$$

$x_{it} = x_{io}[2-\exp(\lambda_{io}t_i)], x_{it}^* = x_{io}^*[2-\exp(\lambda_{io}^*t_i)]$.

After substituting $\lambda_{io} = \alpha_{io} + j\beta_{io}, \lambda_{io}^* = \alpha_{io} - j\beta_{io}, \beta_{io} = \gamma_{io}\alpha_{io}$ at

$\exp(\lambda_{io}t_i) = \exp(\alpha_{io}t_i)[\cos(\beta_{io}t_i) + j\sin(\beta_{io}t_i)], \exp(\lambda_{io}^*t_i) = \exp(\alpha_{io}t_i)[\cos(\beta_{io}t_i) - j\sin(\beta_{io}t_i)]$, we get result

$$H_i^{S_u} = \frac{(\alpha_{io}^2 + \beta_{io}^2)}{\alpha_{io}} \frac{[4-2\exp(\alpha_{io}t_i)\cos(\alpha_{io}\gamma_{io}t_i)+2\gamma_{io}\exp(\alpha_{io}t_i)\sin(\alpha_{io}\gamma_{io}t_i)]^{-1}v_i(t_i)v_i^*(t_i)}{[(2-\exp(\alpha_{io}t_i)\cos(\alpha_{io}\gamma_{io}t_i))^2 + \exp(\alpha_{io}t_i)\sin^2(\alpha_{io}\gamma_{io}t_i))^2]x_{io}x_{io}^*},$$ which brings

$$H_i^{S_u} = H_{io}^{S_u} \times D, D = \frac{2[1-\exp(\alpha_{io}t_i)(\cos(\alpha_{io}\gamma_{io}t_i)+\gamma_{io}\sin(\alpha_{io}\gamma_{io}t_i)]^{-1}}{[(4-2\exp(\alpha_{io}t_i)\cos(\alpha_{io}\gamma_{io}t_i)) + \exp(2\alpha_{io}t_i)]} \text{ at } \frac{v_i^o v_i^{o*}}{x_{io}x_{io}^*} = 4. \quad (3.4)$$

At $t_i \to 0$, $H_i^{S_u} \to H_{io}^{S_u}\delta(t_{io})$, where $\delta(t_{io})$ is delta function, since the multiplier $D \to \infty$.

In this case, the opposite controls are applied simultaneously in moment $t = t_{io}$.

Considering entropy increment at small $\delta t_i$ and let $\alpha_{io}\delta t_i = a$, we get $\Delta S_{i\delta t}^{S_u} \cong a(1+\gamma)3.6 \cong 5.4a, a \cong 0.25$.

The entropy increment at moment $t_i$, count by this formula at $a_o(\gamma_{io} \cong 0.5) \approx 0.75$, is $\Delta S_{it}^{S_u} \approx 1.86a_o$.

Therefore at $\Delta S_{i\delta t}^{S_u} \to \Delta S_{it}^{S_u}$ the increment decreases twice from $\Delta S_{i\delta t}^{S_u} \cong 2.8$ to $\Delta S_{it}^{S_u} \cong 1.4$.

This result takes into account changing $x_i(t_i)x_i^*(t_i)$ regarding $x_{io}x_{io}^*$, as a more general case, but still holds for the simultaneously applied both controls.

Considering ratio $v_i(t_i)v_i^*(t_i) / x_{io}(t_i)x_{io}^*(t_i) = v_i(t_i)v_i^*(t_i) / v_i^o v_i^{o*} \times v_i^o v_i^{o*} / x_{io}(t_i)x_{io}^*(t_i)$ at the controls being proportional to time: $v_i(t_i) = v_i^o c t_{io}$, $v_i^*(t_i) = v_i^{o*}(ct_{io} - \Delta t_i)$, we have $v_i(t_i)v_i^*(t_i) = v_i^o v_i^{o*}(1-\Delta t_i / t_{io})$ and $v_i(t_i)v_i^*(t_i) / x_{io}(t_i)x_{io}^*(t_i) = 4(1-\Delta t_i / t_{io})$, where there is a finite time interval between the applied controls, related to the time interval of applying controls $\Delta t_i / t_{io} \neq 0$.

In this case, the above multiplier is

$$D = \frac{2(1-\Delta t_i / t_{io})([1-\exp(\alpha_{io}t_i)(\cos(\alpha_{io}\gamma_{io}t_i)+\gamma_{io}\sin(\alpha_{io}\gamma_{io}t_i)]^{-1}}{[(4-2\exp(\alpha_{io}t_i)\cos(\alpha_{io}\gamma_{io}t_i)) + \exp(2\alpha_{io}t_i)]}. \quad (3.5)$$

At the same $a \cong 0.25$ for $\Delta t_i / t_{io} = 0.2$, we have $\Delta S_{i\delta t}^{S_u} = 2.24$, while at $\Delta t_i / t_{io} \to 1$, when the shift of the controls approaches the interval of applying controls, the control's interactive entropy moves toward zero: $\Delta S_{it}^{S_u} \to 0$. At $v_i(t_i)v_i^*(t_i) = x_i(t_i)x_i^*(t_i)$, we get the Hamiltonian contribution from the controlled process at the same moment

$$H_i^{S_u} = H_{io}^{S_u} \times \frac{[1-\exp(\alpha_{io}t_i)(\cos(\alpha_{io}\gamma_{io}t_i)+\gamma_{io}\sin(\alpha_{io}\gamma_{io}t_i))]^{-1}}{2[(2-\exp(\alpha_{io}t_i)\cos(\alpha_{io}\gamma_{io}t_i))^2 + \exp(\alpha_{io}t_i)\sin(\alpha_{io}\gamma_{io}t_i))^2]}, \quad (3.6)$$

which is in four times less than that for the controls. Hence, the related entropy contribution $\Delta S_{i\delta t}^{S_x} = 1/4 \Delta S_{i\delta t}^{S_u}$, which at the same $a \cong 0.25$ and $\Delta t_i / t_{io} = 0.2$ brings $\Delta S_{i\delta t}^{S_x} \cong 0.56$, that coincides with its evaluation by invariant $a_o^2(\gamma_{io} \cong 0.5) = 0.5625$. At the controls' starting moment, the process' entropy contribution is $\Delta S_{i\delta t}^{S_{xx}} \cong 0.7$ which we had estimated by $a_o$.

These result shows that each pair of the control contributions in Hamiltonian process sharply increases at the moment of their simultaneous applying and then decreases to $\Delta S_{it}^{S_u} \approx 0.5$, evaluated by $\Delta S_{it}^{S_u} \approx a_o^2$, where imaginary component of the process disappears.

Therefore, the above controls, applied to a non-Hamiltonian process (with only real eigenvalues) hold that evaluation, which specifically corresponds to impulse controls, connecting the dynamic segments at the ends of their optimal processes.

If the controls shift is finite, approaching interval of observation $\delta_{ti} = \Delta t_i / t_{io}$, such control's pair covers actual impulse control, extracting information for the controllable process $\Delta S_{i\delta t}^{S_x} \cong 0.56$, which coincides with evaluation Part1. While each pair of the impulse spends entropy $\approx 1/2 \Delta S_{i\delta t}^{S_x} \cong 0.28$, evaluated by invariant $a \cong 0.25$. We come to previous evaluations of the impulse control entropy's interactive action on the process $\Delta S_{i\delta t}^{S_x} \approx 2a \approx a_o^2$, where $a_o^m = a_o^{m=1}(\gamma_{m=1}^\alpha)^m$ takes into account the process' entropy evaluation $a_o$ by the applied controls pair. Since, simultaneously applied controls lead to delta-function for the entropy Hamiltonian(which is not realistic), the actual double (impulse, or opposite) controls should be applied on a finite time interval.

Let us fund such one that satisfied the observer's mimiax as most acceptable.

If each control carries information $a$, the controls, applied with time shift $\delta_{ti} = \Delta t_i / t_{io}$, hold interactive effect
$$\Delta S_{i\delta t}^{S_u} = 2a(1+\gamma_{it}^2)(1-\delta_{ti}), \tag{3.7}$$
and if the delivered information covers this interactive effect at $\Delta S_{i\delta t}^{S_x} = 2a = \Delta S_{i\delta t}^{S_u}$, then such controls deliver maximal information and parameter dynamic $\gamma_{it}$ depends on interval observation $\delta_{ti}$
$(1+\gamma_{it}^2)(1-\delta_{ti}) = 1$ satisfying Eq.
$$\gamma_{it} = [\delta_{ti}(1-\delta_{ti})^{-1}]^{1/2}, \tag{3.8}$$
which at $\delta_{ti} = 0.2$ brings $\gamma_{io} \cong 0.5$, or vice versa.

Theoretical impulse control leads to $\gamma_{it} \to 0$. Such control brings quantum bit of information with interval of observation $\delta_{ti} \to 0$, this finite minimum approaches to a qubit interval.

At any random observation with a finite classical time interval, the observing random uncertain information is integrating by entropy functional and transformed in certainty path integral.

Classical random information transfers to quantum level at cutting time of observation.

An observer, similar to each chess player, has to integrate multiple observable uncertainties, takes multiple probing actions and integrates them by responding in next move.

While the EF integrates quantum-cut correlation the IPF integrates the observer probing action, and generates the observer inner microdynamic and macrodynamic processes, minimizing a difference of these functionals' observed uncertainty-certainty, which allows observer approaching certainty.

Since the inner opposite controls reflect the observer the impulse control, the information spent on compensating the interactive effect leads to non-symmetry of this inner controls, which for quantum observation loses almost zero symmetry, but for classical observation, asymmetry holds finite size (3.8). (It has shown (Lerner1999), also in Sec.2.6) that the space-time model possesses an odd symmetry of transformation of reflection). While the inner controls, acting on different level of information network (IN), need distinctive time for the IN nodes' cooperation, the related impulse control and the time of observation are the same. This requires changing the inner time scale for observer's inner controls and processes during the observing time.

That means the time scale should be variable:

$$M_m = t_m / \delta_{ti}, \qquad (3.9)$$

where $t_m$ is time interval for $m$ the triplet's node needed for the cooperation.

Assuming $\delta_{ti}$ equals to the interval of cooperation for the very first IN triple $t_{m=1}$, we have

$$M_m = t_m / t_{m=1} = (\gamma^{\alpha}_{m=1})^m, \qquad (3.10)$$

where $\gamma^{\alpha}_{m=1}$ is the IN parameter of multiplication determines by the IN condition of cooperation for the first triplet. Depending on fixed $\gamma_{it} = \gamma$, this parameter determines the following invariant of cooperative dynamics: $\gamma^{\alpha}_{m=1}(\gamma \to 0) = 4.794, \gamma^{\alpha}_{m=1}(\gamma = 0.5) = 3.895, \gamma^{\alpha}_{m=1}(\gamma = 0.8) = 3.3$.

Obeying relation (3.8a) involves multiplication of intensity of the observer's processes (shortening their time scale), characterized by amplitude of information flow $I_m = \alpha^m_{i\tau}$, depending on the IN $m$-th level of cooperation, whose $m$ node concentrates the multiplied density of information $I_m = I_{mo}^{m=1}(\gamma^{\alpha}_{m=1})^m$ regarding that for the IN first triple $I_{mo}^{m=1}$.

The IN density measure quality of the IN hierarchical level enfolding in that node, which applied to the IN highest level portrays the observer's intellectuality (Lerner 2012,20113). This means, the related observed information should also enfold that information density, or information, concentrating in the observation, embraces the quality needed for that hierarchical level.

The observer enables to extract such information and enfold it in the IN higher level (increasing the quality), automatically multiples its time scale of intensity of inner information processes.

The recent study on neuronal mechanism of learning (Reuveni, Saar, Barkai 2013), supports this theoretical results, demonstrating that the events amplitude, proceeding in neuronal information network during the learning, is multiplied by average factor 2.3-2.5 on each network level through doubling the strength of all synapses in the level's subgroup of cells, as a response of already existing memory.

The controllable micro-macrodynamics and IN, through the minimax principle, the hidden uncertainty, which observer cuts off from correlations, reaches equivalence with the certainty, creating mathematical, informational and logical self-consistence.

The recurring integration of requested information replenishes the IN hierarchy, increasing its subsequent level's quality and density via strengthening of information processes and co-operations, which concurrently are memorized and enhanced in the IN logic.

## 2.4. *A time-space information process, its information structure and the boundaries.*
## *Forming Information Network (IN) of the process with the IN boundaries*

We model a space distributed random process by solution of a controlled stochastic Eqs (Part1), where both drift and diffusion functions depend parametrically on a non-random vector of geometrical coordinates (Lerner 2008b, 2010). We have defined the entropy functional for this model and found the solution of the VP problem with related IPF.

In the space-time consolidated model, the required diagonal zing of the dynamic operator proceeds under sequential rotations of its initial spectrum (Fig.2.2), following from VP.

Fig.2.3 shows the equalization of the model's eigenvalues for the corresponding eigenvectors during the optimal movement with the triplet node's formation at the localities of the triple cones vertexes' intersections. Fig.2.4 illustrates the structure of a cone, formed by a rotating spiral.

A space operator, performing the rotation, is formed using the eigenvectors of models' dynamic operator (Secs 1.6 and 2.1). Such a space movement, directed toward diagonal zing of the dynamic operator, is an attribute of the space consolidation process, which leads to both arranging of the ranged sequence in the triplet eigenvalues sequence and providing their consolidation in the IN's triplets structures.

The rotating operator retains a symmetry of the transformation only within each discrete interval when VP is satisfied.

A common symmetry of transformation, applied to all elements of the *n*-dimensional operator, following from VP, connects information dynamics and geometry (Lerner 2005).

By the end of discrete interval, the elements of operator complete diagonal zing that allows the rotating eigenfunctions (as the eigenvalues) proceed toward their subsequent cooperation in triplets.

The informational dynamics in time-space, described by a sequence of the IPF space distributed extremal segments, form spiral trajectories, located on a conic surface, while each segment represents a three-dimensional extremal.

The trajectories of each conjugated dynamics compose an opposite directional double spirals (Fig.2.5) rotating on the surfaces of the same cone. On the cone's vertex, the opposite trajectories join together holding their equal real eigenvalues. Such optimal double spiral geometry follows from each pair of the model's Hamilton Eqs, distributed in space, whose conjugated space solutions-waves (Sec.1.11) decrease exponentially, merging at the cone's vertex.

Each triple of the pair conjugated eigenvalues forms three-dimensional conjugated components of its eigenvector. Rotation of these eigenvectors (in the direction of their consolidation, required by the VP) takes a shape of double-spiral structure (Fig.2.5b,c) with opposite spirals' rotations.

An intersection (juncture) of these spirals (Fig.2.5d) corresponds to the UR zone, where each triplet's node emerges. A detail example of forming three rotating eigenvalues for a three-dimensional dynamics operator is presented in (Lerner 2010a, p. 158-163).

Both the directional movement of each triplet's eigenfunction and the consolidation of their components (eigenvalues) proceed under the controls (Secs.1.6,2.3) acting on a window between cooperating segments.

The implementation of the IPF minimax principle leads to a sequential assembling of a manifold of the process' extremals (Fig.2.3) in elementary binary units (doublets) and then in triplets, producing a spectrum of coherent frequencies. The manifold of the extremal segments, cooperating in the triplet's optimal assemble, are shaped by a sequence of the spirals conic structures, whose cones' vertexes form the nodes of an information network (IN) (2.4). In the hierarchy of the IN nodes, the IN-accumulated information is conserved in the invariant form.

Each IN node is assembled during the space-time dynamics within the considered UR zone with volume $V_c$ of already formed triple zone, whose volume's increment $\partial V$ during time $t^3$, is evaluated by size of UR (2.9):

$$\delta V = V_c \delta t^3, \delta V^* = (\delta V / V_c t^3) = \varepsilon(\gamma)^3. \tag{4.1}$$

This relative form $\partial V^*$ estimates a size of an IN node.

An external information, associated with current process' data, is delivered at the punched localities of the IN external surface (Sec.2.6), which shape the UR.

Considering $\varepsilon_k$ as a relative average quadratic error, limiting cooperation, we will find such initial IN information speed $\alpha_{i=1,o}$, which restricts the minimal speed of cooperation (2.9a).

Since with growing number of cooperating triplets $m$ each triplet information speed $\alpha_{i+1o}^m(t_k^{i+2})$ decreases, we find first such $m_o$ for which cooperative speed reaches its minimum (2.9a) at $\alpha_{i+2o}^m(t_k^{i+2}) = \alpha_{i+2o}^{m=1}(t_k^{i+2})/m_o$, $C_{mo} = c_{ico}/\alpha_{i+2o}^{m=1}(t_k^{i+2}) \cong (2\mathbf{a}_i/\mathbf{a}_{io})/m_o \cong 0.66/m_o$, which also restricts the IN relative $C_{mo}$ speed of encoding symbols.

The time intervals $t_k^{i+2} = t_k^m$ increase with growing triplet number $m$, so their ratio $t_k^{m=1}/t_k^m$ decreases, reaching admissible minimum $t_k^{m=1}/t_k^{m_o} = \partial t_k/t_k = \varepsilon_k$, corresponding ratio $\alpha_{i+2o}^m(t_k^{i+2})/\alpha_{i+2o}^{m=1}(t_k^{i+2})$ and equals to $C_{mo}$. We get $m_o = 0.66/\varepsilon_k$, or at $\varepsilon_k \cong 0.1816$, the maximal admissible IN level, restricting the encoding speed is $m_o \cong 8.5$.

This limits maximal admissible dimension of cooperating processes $n_o \cong 2m_o \cong 17$. This limitation is closed to the evaluation of neuron net required for human retina (Comment 2.3).

Using $\alpha_{i=1,o}^{m=1} = \gamma_i^\alpha \alpha_{i+2,o}^{m=1}$ we can find $\alpha_{i=1,o}^{m=1} = \gamma_i^\alpha c_{ico} m_o / 0.66$ which at $\gamma_i^\alpha \cong 3.9, \gamma = 0.5$ is an equivalent of increasing encoding speed by $\alpha_{i=1,o}^{m=1} \cong 50 c_{ico}$.

Applying Prop. 2.1 and the IN hierarchy, we come to the following specific results.

<u>Corollary 2.4a.</u>

(1). Each process' eigenvalue $\{\alpha_{io}\}$, identified during the time course, includes the *current* process' data, coming from each punched locality DP(*i*) of the random process, where the identification takes place;

(2). Influx of the data starts with the identified maximal eigenvalue $\alpha_{1o}$ (at the segment with minimal time interval $t_k^1$), continues consequently with the time course, and ends with the identified minimal $\alpha_{nk}$ (at maximal time interval $t_k^n$);

(3). An hierarchy of the IN nodes, originated from maximal $\alpha_{1o}$, collects a current process information, starting with the process' minimal time interval and ending with minimal $\alpha_{nk}$ and maximal time interval $t_k^n$ - at the IN *final node* (modeling the IN highest hierarchical level), which collects a *total* amount of data coming from all previous nodes;

(4). The IN information, acquired by each node, satisfies the invariant relations, following from (Sec 2.1);

(5). The controls, implemented the above relations, progressively increase the number of cooperating IN nodes, which are enclosed into its final node;

(6). The information, transformed from each IN previous to the following triplet, has an increasing value, because each following triplet encapsulates and encloses the total information from all previous triplets;

(7). The node location within the IN hierarchy determines the *value* of information encapsulated into this node, which is estimated by the ratio $\gamma_m^\alpha(\gamma_m) = \alpha_{om-2}/\alpha_{om}$ for the $m$-th triplet number, where $\alpha_{om}$ is triplet's starting eigenvalue;

(8). Since each following IN node encloses the previous triplet's information, the IN's final node accumulates all its information, enclosing information from all model's spectrum.

(9). Because each IN's node accumulates information contribution (2.3), this entropy speed decreases under each space-time cooperation. •

As a result, each triplet, spending on cooperation additional information $\mathbf{a}_o^2(\gamma)$, gets its stable cooperative structure, whose IN's node accumulates information of growing quality.

### *2.4a. The IN information code*

In the cooperative process of triplet formation, three starting step-up controls initiate dynamics for each segment, and then each of the following two impulses' controls change the initial dynamics these two segments to reach the equalization of their ending eigenvalues with the ending eigenvalue of a third segment, whose dynamics had started simultaneously with the initial dynamics of the first two segments. Since actions of these total seven step-up controls proceed during the time interval of the third segment's dynamics, their seven digits *sequentially appear within this time interval.* By the end of this time interval, the two subsequently applied impulse controls connect all three segments with their equal eigenvalues in a joint triple cooperative unit, during the time interval of these segments' windows. Hence, at the widows sequentially appear 7+4 digits of the total stepwise controls actions, which generate the triplet during the time interval of the third' segment's dynamics and its window time duration (that embraces both windows of two attached segments).

Since these 11 digits produce 5.5 digital impulses, while each such impulse carries information in 0.5 Nats, these controls bring information in 2.75 Nats or approximately four bits.

Specifically, using the mode's invariant information measure of elementary information, hidden by a pair correlation (Sec.2.3) and delivered through a stepwise control $\mathbf{a}(\gamma) \cong 0.25$ Nats, we evaluate the code information by $11\mathbf{a}(\gamma)$ Nats, or $\cong 2.75$ Nats – four bits.

This invariant measure is not based on the elementary information measure of impulse.

These discrete units of the triplet's controls form a *triplet's code* that governs formation of the optimal triplet.

Information, carried by all triplets' segments up to their cooperation in the triplet's ending node, is enclosed into a single unit of $\mathbf{a}_o(\gamma) \cong 0.75 Nat \cong 1 bit$, which supposed to accumulate the triplet code.

This means that the four digits of the triplet's code could be encoded by a single digit, which is conserved in the IN node. Because the last segment of that triplet (with information 1bit) forms the first segment of the next forming triplet, it brings 1 bit information to the emerging triplet's information code of 3 bits, making a total 4 bit, where 1 bit from the previous triplet carries out all its information that had been encoded.

In fact, three bits (from the total triplet's four bits) encode each triplet's joint node by a single bit, while one bit (of the four), corresponding to invariant $\mathbf{a}_o(\gamma) \cong 2\mathbf{a}(\gamma)$, is transferred to a next triplet.

This information $2\mathbf{a}(\gamma)$ is necessary to provide two stepwise controls, starting the following two macrodynamic process on the next triplet.

Four digits-symbols, produced at a triplet's window, have information $2\mathbf{a}_o^2(\gamma) \cong 1 Nat$, which corresponds to the number of equal probable combinations with these four symbols $q = exp1 \cong 3$.

This means, each of four symbols could appear in the triple with probability 1/3.

Consequently, when the control closes the window, any of this sequence of symbols could be remembered as a triplet code only with probability 1/3.

A code with such uncertainty holds this probability of errors (mutation), allowing to encode a new substance in any alternative code sequence (in their potential combinations).

Fourth digit-symbols of the code could serve for the code protection, as a marker of a correct code sequence: a wrong place of this marker would detect the code error.

Such a single digit's unit provides only one–third of the code protection, leaving two of the probable three code's combinations be allowable.

Each triplet's information code should be generated externally during the triplet's formation in coordination with the cooperative dynamics.

From other side, the code is both an attribute and a result of the triplet's cooperative dynamics.

The code depends first on the initial information speeds of each its three segments, their location within the IN, and also from the current information, obtained from the segments' window during the cooperative dynamics (that could deliver a code's external variation-mutations).

A given code could generate the triplet's information structure, if an information hardware, operating the system, synthesizes the code's software information via implementing the code's dynamic operations.

Each code of the current triplet encodes not only the structure of that triplet, but also enfolds a code of all previous IN's nodes. Such a code accumulates and encodes the information values encapsulated into these upper nodes.

Thus, a sequence of the successively enclosed triplet-nodes, represented by discrete control logic, creates the IN code with a three digits from each triple segments and a forth digit from the control that binds the segments.

The code serves as the IN virtual communication language and an algorithm of minimal program to design the IN.

The optimal control (Secs.1.6,1.11), formed by copying and doubling of an identified state, produces two opposite directional dynamic trajectories, starting each with a copying state.

In a space-time, these trajectories form the double spiral structure, whose parts cooperate at each triple cooperation, possessing a triplet code of four symbols (by two from each spiral).

The code is located between these spirals at each UR zone of the triplet's formation (Fig.2.5d).

The conic structures (Figs.2.1, 2.4), generating the UR cells, form an information background locations for the IN. At a given cells' logic, the sequence of cells' blocks forming the IN nodes, can be built. For such an IN, both $\mathbf{a}_m(\gamma)$ and $\mathbf{a}_{mo}(\gamma)$ are the $m$ space distributed node's invariants, preserving their scalar components ($\mathbf{a}(\gamma)(\gamma)$ and $\mathbf{a}_o(\gamma)$) for a fixed $\gamma$, and depending geometrically on the relative shifts of angle $\psi_m(\gamma_m^\alpha)$ (Fig.2.3) and the angle starting position.

As a result, the optimal IN code's geometry has the double spiral (helix) triplet structure (DSS) (Fig.2.5), which is sequentially enclosed in the IN nodes, while its final node enfolds a total DSS. Decoding this node allows the reconstruction of both the IN dynamics and topology.

The IN information geometry holds the node's cooperative binding functions and an asymmetry of triplet's structures. In the DSS *information geometry,* these binding functions are encoded, in addition to the encoded nodes' dynamic information. The DSS specifics depend on the structure of the entropy integral (EF) functions' drift and diffusion in Eqs (1.2.6)(also in Sec.2.3).

The IN is built by the spectrum of the model's eigenvalues, while implementing the VP.

Such a spectrum, satisfying the VP, is able of a self-creation from its IN ending node, which encloses all spectrum information, measured by the EF-IPF functional.

*2.5. Connection to Shannon's information theory. Encoding the initial information process*

Information of observed random process, encoded in the DSS, can be expressed in a sequence of an elementary code-word lengths in a standard format of information code for communications.

Considering a set of discrete states $\tilde{x}(\tau^o) = \{(\tilde{x}_i(\tau_k^o)\}, i=1,...,n; k=1,...,m$ at each fixed moments $\tau_k^o$ along $n$-dimensional random process, and using definition of the entropy functional (1.2.6), we get the conditional entropy *function* for the conditional probabilities (corresponding to (1.1.1)) at *all* moments $\tau_k^o$ at each process dimension $S_{\tau_k^o}^i$ *and* for the whole process $S_{\tau_k^o}$ accordingly:

$$S_{\tau_k^o}^i = -\sum_{k=1}^m p_k[\tilde{x}_i(\tau_k^o)]\ln p_k[\tilde{x}_i(\tau_k^o)], S_{\tau_k^o} = \sum_{i=1}^n S_{\tau_k^o}^i, \tag{5.1}$$

which *coincides with the Shannon entropy* for each probability distribution $p_k[\tilde{x}_i(\tau_k^o)]$, measured at each fixed $\tilde{x}_i(\tau_k^o)$. Function (5.1) holds all characteristics of Shannon's entropy, following from the initial Markov process and its additive functional for these *states*.

For the comparison, the controllable IPF entropy (information), measured at the related discrete punched localities (at the widows):
$\tilde{x}(\tau) = \{\tilde{x}(\tau_k)\}, k=1,...,m$

$$\tilde{S}_{\tau m}^i = \sum_{k=1}^m \Delta S_k[\tilde{x}(\tau_k)], \tag{5.2}$$

(where $\Delta S_k[\tilde{x}(\tau_k)]$ is the entropy at each $\tau_k, k=1,...m$) ;which is *distinctive* from the entropy $S_{\tau_k^o}^i$ (5.1), because:

(1)-The IPF entropy $\Delta S_k[\tilde{x}(\tau_k)]$ holds the macrostates' ($x_i(\tau_k-o), x_i(\tau_k), x_i(\tau_k+o)$) *connection* through the punched locality (performed by applying the two step-wise controls), while the EF binds all random states, including the punched localities;

(2)-The distribution $p_k = p_k[\tilde{x}(\tau_k)]$ is selected by variation conditions (1.5.1, 1.5.1a) (applied to (1.3.7a)), as an extremal probability distribution, where a macrostate $x(\tau_k)$ estimates random state $\tilde{x}(\tau_k)$ with a maximum probability $p_k = p_k[\tilde{x}(\tau_k)]$.

At this "sharp" probability maximum, the information entropy (1.3.7) reaches its local maximum at $\tau_k$-locality, which, for the variation problem (1.5.1), is associated with turning the constraint (1.5.10) off by the controls, switching the model to the random process.

Selecting these states (with an aid of the dynamic model's control) allows an *optimal discrete filtration* of random process at all moments $\tau_k, k=1,...m$, where the macromodel is identified and external information is delivered.

Thus, $x(\tau_k)$ emerges as the *most informative* state's evaluation of the random process at the $\tau_k$-locality.

Even though the model identifies a sequence of the most probable states, representing the most probable trajectory of the diffusion process, the IPF minimizes the EF entropy functional, defined on a *whole* diffusion process.

Therefore, the IPF implements the optimal process' functional information measure.

Each of the entropy $\Delta S_k[x_i(\tau_k-o)]$ measures the process segment's undivided information, and the entropy $\Delta S_k[x_i(\tau_k)]$ delivers *additional* information, compared to the traditional Shannon entropies, which measure the process' sates at the related discrete moments.

Here we evaluate information contribution for each segment by the segments' $\tau_k$-locality $\Delta S_k[x_i(\tau_k)]$. Applying the invariant's information measure to this total contribution $3\mathbf{a}_i(\gamma_i) \cong \mathbf{a}_{io}(\gamma_i)$ (which includes all information delivered with the impulse control), we get the total process $\tilde{S}_{\tau m}^i$ *estimation* by the sum of the invariants, which count both the inner segment's and control inter-segment's information:

$$\tilde{S}_{\tau m}^i \cong \sum_{k=1}^{m} \mathbf{a}_{ok}(\gamma_k), \tilde{S}_\tau = \sum_{i=1}^{n} \tilde{S}_{\tau m}^i, \qquad (5.3)$$

where $m$ is the number of the segments, $n$ is the model dimension (assuming each segment has a single $\tau_k$-locality).

Therefore, to *predict* each $\tau_k$- locality, where $\Delta S_k[\tilde{x}(\tau_k)]$ should be measured, we need only each invariant $\mathbf{a}_{ok}$ which estimates the IPF entropy with a maximal process' probability.

Knowing *this* entropy allows encoding the *random process* using the Shannon formula for an average optimal code-word length:

$$l_c \geq \tilde{S}_\tau^o / \ln D, \qquad (5.4)$$

where $D$ is the number of letters of the code's alphabet, which encodes $\tilde{S}_\tau^o$ (1.2.4).

An elementary code-word to encode the process' segment is

$$l_{cs} \geq \mathbf{a}_{ok}(\gamma_k)[\text{bit}]/\log_2 D_o, \qquad (5.5)$$

where $D_o$ is a segment's code alphabet, which implements the macrostate connections.

At $\mathbf{a}_{ok}(\gamma_k \to 0) \cong 1 bit$, $D_o = 2$, we get $l_{cs} \geq 1$, or a bit per the encoding letter.

Therefore, invariant $\mathbf{a}_{ok}$ allows us both to encode the *process* using (5.4), (5.5) including both the segments' and the between segments' information contributions *without* counting each related entropy (5.2), and to compress each segment's random information to $\mathbf{a}_{ok}$ bits. With such optimal invariant, the quantity of encoding information for each segment will be constant, while a width of the encoding impulse $\delta\tau_k^i$ (1.8.14a) depends on the quantity of external information.

Since interval between these impulses $t_k^i$, measured by invariant $\mathbf{a}_{ok}$, will be also constant, such encoding is described by a varied pulse-width modulation. At any fixed $\mathbf{a}_k \neq \mathbf{a}_{ok}$, both interval $t_k^i$ and pulse-width will be different, and the encoding is described by a pulse-amplitude modulation with a varied time. At a fixed external information, the various $\delta\tau_k^i$ will generate different $t_k^i$, which is described by pulse-time modulation (Figs1.3a-c, Part1).

The assigned code also encodes and evaluates both constraint and controls' actions.

Under the constraint, *each stochastic equation* (1.2.1) with specific functions of the drift-vector and diffusion components *encloses a potential number of the considered discrete intervals and their sequential logics.*

The selection of both the discrete intervals and their numbers is *not arbitrary*, as it is in the known methods (Shannon 1949, Stratonovich 1975), since each specific selection, following from the VP, is applied to the *whole* process.

The developed procedure (Secs.1.10-1.11)(Lerner 2008b, 2010a) allows building the *process hierarchical information network* (IN), which, along with encoding a sequence of the process extremals in a *real* time, also ranges the segments by the values of their local entropies–invariants.

The IN building includes the eigenvector's ranging and their aggregation in the elementary cooperative units (triplets), where the IN accumulated information is conserved in the invariant form. Each current IN's triplet, evaluated by the invariant quantity of information, can be encoded in its triplet's code. Thus, each process (1.2.1) with its space distributed hierarchical (IN) structure, which is built during the process' real time, can be encoded by the IN's individual information code. Each current IN's triplet, evaluated by quantity of information $\mathbf{a}_i(\gamma_i) + \mathbf{a}_{io}(\gamma_i) = \tilde{S}_\tau^o$, can be encoded in its DSS triplet's code.

Therefore, each information process (Sec. 2.1) with its space distributed hierarchical (IN) structure, which is built during the process' real time, can be encoded by the IN's individual DSS code.

The cooperative IN's specific consists of revealing the process *interactive dynamics*, which emerge by the dynamic binding of the process' information and the segments' macrodynamics into a *dynamic information system with the system's code hierarchy.*

Information, collected by the Entropy Functional (EF), is transformed to the VP selected portions of Information Path Functional (IPF), which are separated by the windows between them. The windows give access of new information flow that can renovate each subsequent the IPF portion.

The IPF provides optimal intervals for both measuring incoming information and its consequent accumulation. This includes local maximums of the incoming information, which are accumulated by the IPF portions, while their connection in a chain minimizes the total information.

The implementation of the VP minimum for the EF leads to assembling of the above chain's portions in the information dynamic network (IN). The IN hierarchical tree is formed by sequential consolidation of each three IPF portions into the IN triplet's nodes, which are successively enclosed up to formation of a final IN node, which accumulates all IN enclosed information.

The EF-IPF approach converts a random process' *uncertainty* into the information dynamic process' *certainty*, with its code and the IN. This information formations implement the principle of minimum for the *potential information paths*, connecting observed uncertainty with subsequent formed optimal information-certainty during the information *conversion process' time course*.

This minimum principle is a particular information form of the *fundamental minimum principle* in Physics (Landau and Lifshitz, 1965).

*Example of the IN* information hierarchy, enclosed in the IN nodes, computed by developed software (Lerner 2013), is shown on Figs.2.4.

## 2.6. The IN's external surface and its geometrical structure

The model's space-time distributed dynamics self-develops a topological structure of information geometry (Lerner 2010a), which is limited by the information structure of the IN nodes' hierarchy.

This information boarding surface is formed by a sequence of the IN's nodes surfaces, emerged at the end of rotated spiral time-space trajectory, which are located on the cones' surfaces (Fig.2.3) along each local extremal movement.

Each of the node external surface is measured by the size of related UR (4.1, 2.6a), where the segments' dynamics interact with environment.

The UR external surface with area $\varepsilon^2(\gamma)$ forms a curved cell with area $f_o \cong \varepsilon^2(\gamma)$, whose cell's side is measured by $\varepsilon(\gamma)$. This cell unifies information other elementary cells from the IN hierarchy by enclosing information units of the triplet's code, whose information is delivered from the interactive dynamics. Since the triplet code consists of four bits, the UR cells are made of four elementary cells (each 1 bit) with area of each elementary cell $f_c^o \cong \varepsilon^2(\gamma)/4$. After the cooperation in a node, these cells are compressed in a triplet's final joint cell (encoding 1 bit), which, moving to the following triplet, transforms its 1 bit information to the ending triplet's code.

In the space-time consolidated model, the required diagonal zing of the dynamic operator proceeds under rotations its initial process' spectrum (Sec.2.1), following from VP. Such a space rotation completes the operator diagonal zing by the end of the discrete interval, which is accompanied by the joint macrocells' rotation toward a subsequent triplet's cooperation.

During this space-time movement in the optimal process, the IN cellular external surface automatically emerge. This means, the model's hierarchical information structure creates its own geometrical boundary during its self-formation along the process' time course.

As a result, the IN external a hierarchy of discrete cellular structures composes information geometry, whose cells' geometry enfolds the code of each IN node.

The model interacts with an environment through the IN external surface of this hierarchical cellular geometry, which delivers external information.

Interaction could be in any node-cell's location on the surface.

Since concentration of the IN cells' information quality (values) depends on the node-cell location on the surface, the interactive dynamics have a selective reflection along the IN's hierarchical cellular surface, decreasing its highest information value from the IN final node to the lowest information value for a starting node.

The reflected information is shaped according to these values, forming of a related IN structure having space-time images. The external surface's finite square with its cellular geometry imposes a limit on the model's interactive information, restricting the flows of information and the reflected space images information structure.

## 2.6a. Analysis of the surface structure formed by the IN cellular geometry: the cell-space area, its curvature and rotation dynamics

The surface area (Fig.2.5), defined by function of its current radius $y$ in the form $F = \pm \pi y^2$ (Lerner 2010), is shaped by rotation of a hyperbolic function $y = a/x$ (6.1) ( Fig. 2.5a) around axis $0 - x$.

For the considered IN with $m$ triplet's nodes, such function corresponds to the model's invariant relation in the form

$$y = \mathbf{a}(\gamma)/\alpha_m(t_m) = t_m, \tag{6.1a}$$

where $\alpha_m(t_m) = \alpha_m$ is the node eigenvalue at discrete moment $t_m$ of creating $m$-th triplet, $\mathbf{a}(\gamma) = \alpha_m t_m$ is the model invariant depending on parameter $\gamma$, which is fixed for each current model.

The eigenvalues of a nearest triplets' area are connected by the ratio

$$(\alpha_{m-1}/\alpha_m)^2 = (\gamma_{m-1}^\alpha)^2, \tag{6.2}$$

where for optimal IN $\gamma_{m-1}^\alpha = \gamma_m^\alpha = \gamma_2^\alpha$. The related surface area:

$$F_m = \pi \mathbf{a}(\gamma)^2 / \alpha_m^2(t_m) = \pi t_m^2. \tag{6.3}$$

Location of each triplet's node forms a spot with an elementary space area $\delta F_m(\delta t_m) = \pi \delta t_m^2$, which, relatively to $F_m$, acquires an invariant form:

$$f_o = \delta F_m(\delta t_m)/F_m = \delta t_m^2/t_m^2 = \varepsilon_m^2(\gamma). \tag{6.3a}$$

This node's spot $f_o$, consists of the number of information cells $m_c$ with cell space area $f_o^c$ each, which allows measuring $f_o$ by the node's code:

$$m_c f_o^c = f_o. \tag{6.3b}$$

Let us find number of the elementary space areas, occupying total space areas, which are formed by the end of a macrodynamic process $T_m$:

$$F(T_m) = \pi T_m^2. \tag{6.3c}$$

Following (6.3), (6.3a-c), we get the number of cells, formed by the process' end:

$$N_e = F(T_m)/\delta F_m(\delta t_m) = T_m^2/\delta t_m^2 = \varepsilon_m^2(T_m^2/t_m^2)(\delta t_m^2/t_m^2) = \gamma_2^{2m}\varepsilon_m^2, \tag{6.4}$$

at $\gamma_2^{2m} = (\gamma_2^\alpha)^{2m}$, where

$$N_e = \gamma_2^{2m} f_o = \gamma_2^{2m} m_c f_o^c \tag{6.5}$$

is total number of the process' cells-codes (bits).

The IN's information density $m_{cm}^c$ of code $m_c$, concentrated in this cell, defined by $m_{cm}^c = m_c \gamma_2^m$, is growing with increase of the node's number $m$.

Using total IN code's information density $m_{cm}^{Nc} = m_c \gamma_2^{3m}$, we get a total cells-codes numbers needed to encodes all process, including the very first triplet with its $f_o^c$ and $\gamma_2^\alpha$:

$$N_e^c = m_{cm}^{Nc} f_o^c. \tag{6.6}$$

The curvature in an information space is defined (Lerner 2006a) by an increment of information speed on an instant of geodesic line in Riemann space $ds$ being relative to information speed on this instant: $K_\alpha^i = -3\dot{\alpha}_o \alpha_o^{-1}, \dot{\alpha}_o = d\alpha_o/ds,$ (6.7) where the increment holds information acceleration $d\alpha_o/dt$ multiplied on inverse space speed $c = ds/dt$ along geodesic line: $\dot{\alpha}_o = d\alpha_o/ds = c^{-1}d\alpha_o/dt, c = ds/dt$, which brings

$$K_\alpha^i = -3(c\alpha_o)^{-1}d\alpha_o/dt \tag{6.7a}$$

Since the curvature is defined on the cooperating three dimensional geodesic lines (three extremal segments), relation (6.7a) measures emergence of the information curvature resulting from non-zero of the relative acceleration at finite space speed. This cooperative information curvature grows with increase of the acceleration, related to information speed, and decrease of the space speed.

In the IN, each of such cooperation creates a triplet, for which curvature of a $m$-th triplet is
$$K_\alpha^m = -3\dot\alpha_m \alpha_m^{-1} \tag{6.7b}$$

Or in the form (6.7a) at a fixed speed $c$:
$$K_\alpha^m = -3\alpha_m \Delta\alpha_m / \Delta t_m \text{ at } \Delta\alpha_m = (\alpha_m - \alpha_{m-1}), \Delta\alpha_m / \alpha_m = (1 - \alpha_{m-1}/\alpha_m) = 1 - \gamma_m^\alpha \text{ and}$$
$$\Delta t_m = (t_m - t_{m-1}), \Delta t_m / t_m = (1 - t_{m-1}/t_m) = 1 - (\gamma_m^\alpha)^{-1} \text{ with } \mathbf{a}(\gamma_m^\alpha) = \alpha_m t_m,$$

the triplet curvature acquires the form
$$K_\alpha^m = 3\alpha_m^3 \gamma_m^\alpha / \mathbf{a}(\gamma), \quad K_\alpha^{m-1} = 3\alpha_{m-1}^3 \gamma_{m-1}^\alpha / \mathbf{a}(\gamma), \tag{6.8}$$

where at $\mathbf{a}(\gamma_m^\alpha) = \mathbf{a}(\gamma_{m-1}^\alpha)) = inv$ and $\gamma_{m-1}^\alpha = \gamma_m^\alpha = \gamma_{m=1}^\alpha$, we come to
$$K_\alpha^m / K_\alpha^{m-1} = (\gamma_{m-1}^\alpha)^{-3}. \tag{6.8a}$$

An increment of curvature (6.7b), related to the spot area, during the time interval $\Delta t_m$ of forming this spot area is $K^{\delta t} = \delta\alpha_m(t_m)/\alpha_m$, which at $\delta\alpha_m(t_m) = -3/2\mathbf{a}(\gamma)\delta t_m / t_m^2$, $t_m^2 = F_m / \pi$, $\mathbf{a}(\gamma) = \alpha_m t_m$ acquires form
$$K^{\delta t} = -3/2 \delta t_m / t_m^3 = -3\pi/2 \varepsilon_m^{1/2} / F_m. \tag{6.8b}$$

This means, the curvature is declining with growing space area $F_m$ and vice versa.

For example, when a triplet's space area $\varepsilon_m$ declines in four time, being concentrated in a single node, we get the increase of curvature also in four times:
$$K_{m\varepsilon}^{\delta t} / K_m^{\delta t} = 4 \text{ at } F_m = \pi\varepsilon_m, \quad F_{m\varepsilon} = 1/4\pi\varepsilon_m. \tag{6.9}$$

So, the initial triplet's area is curved in $4(\gamma_2^\alpha)^m$ times, compared with the area of the IN final node, since more information is concentrated in each following node.

Otherwise, the curvature, being multipled on the square area, brings an invariant:
$$K_m^F = K^{\delta t} F_m = -3\pi/2 \varepsilon_m^{1/2} = inv(\gamma), \tag{6.9a}$$

depending on dynamic parameter of the dynamics $\gamma$, measured by the initial information.

Using relation (6.2), (6.3), we get the ratio of a nearest square areas in the form
$$F_\alpha^m / F_\alpha^{m-1} = (\alpha_{m-1}/\alpha_m)^2 = (\gamma_{m-1}^\alpha)^2. \tag{6.10}$$

Considering volumes of the nearest IN nodes, we come to
$$V_\alpha^m / V_\alpha^{m-1} = (\alpha_{m-1}/\alpha_m)^3 = (\gamma_{m-1}^\alpha)^3. \tag{6.10a}$$

The same relations follow from the details of the IN geometrical structure (Lerner 2010a).

Let us applying the segments' time intervals $(t_5 - t_3), (t_5 - t_4), (t_7 - t_5)$ at forming two nearest triplets (for second and third in the IN hierarchy), for counting the volumes during these triplets' creation:
$$V^{m=2} = V(t_5) = 2V_c 3(t_5 - t_3)^3 + (t_5 - t_4), V^{m=3} = V(t_7) = 2V_c 3(t_7 - t_5)^3 + (t_7 - t_6), \tag{6.11}$$

where $V_c$ is a starting volume for this model. The ratio of these volumes:
$$V^{m=3}/V^{m=2} = [3t_5^3(t_7/t_5 - 1)^3 + t_6(t_7/t_6 - 1)]/[3t_3^3(t_5/t_3 - 1)^3 + t_4(t_5/t_4 - 1)]$$
$$= [3t_5^3(\gamma_2^\alpha - 1)^3 + t_6(\gamma_2^\alpha/\gamma_1^\alpha - 1)]/[3t_3^3(\gamma_2^\alpha - 1)^3 + t_4(\gamma_2^\alpha/\gamma_1^\alpha - 1)]$$

at the equal basic ratios $\gamma_2^\alpha / \gamma_1^\alpha \approx 1$, gets the form

$$V^{m=3} / V^{m=2} \cong (\gamma_2^\alpha)^3 . \tag{6.11a}$$

The related square areas:

$$F^{m=2} = F(t_5) = 2F_c 3(t_5 - t_3)^2 + (t_5 - t_4), \quad F^{m=3} = F(t_7) = 2F_c 3(t_7 - t_5)^2 + (t_7 - t_6)$$

and their ratio at the same $\gamma_2^\alpha / \gamma_1^\alpha \approx 1$, acquire view

$$F^{m=3} / F^{m=2} \cong (\gamma_2^\alpha)^2 . \tag{6.12}$$

This confirms the equivalence for each of the above ratios, counted for:
(1)- the space area (Figs.2.4,2.5), following from the rotation of a hyperbola in equation (6.1), and for
(2)-each triplet's geometrical structure, evolving in their macrodynamic processes.

The geometrical transformations, leading to a triplet's formation, are nonsymmetrical: that includes changing the order of symmetry at sequence of transformations, involving a right directional and left directional rotations of spirals on the conic surfaces (Figs.2.1, 2.4,2.5).

As a difference from symmetrical transformations (at forming crystalloid structures, which hold an even order of symmetry), each triplet's structure holds an odd symmetry order $\Pi_c$, determined by the equations for right directional and left directional rotations accordingly:

$$\Pi_c = \frac{2\pi}{\pi/2 \pm (\pi/4 - \arcsin(2k)^{-1})}, \quad \Pi'_c = \frac{2\pi}{\pi/4 - \arcsin(2k)^{-1}}, \tag{6.13}$$

where at the angle's parameter $k = 1$ we have
$\Pi_c(\psi) \cong 3, \Pi_c(-\psi) \cong 7, \Pi'_c(\psi) = 9$, at the same angle of rotation $\psi$ of the local coordinate systems, which is preserved at the symmetric transformation of local coordinates and it possesses the symmetry of a *reflection*.

At such symmetrical transformations, the positions of the triplet's local coordinate axes in the space are *not* repeated precisely, and any other symmetrical transformations cannot bring the above axes to an equivalent position. This means, each cell's surface area is *non symmetrical*.

We get also a relative curvature for each relative area

$$(K_m^F / F_m) / K_{m-1}^F / F_{m-1}) = (\gamma_{m-1}^\alpha)^{-5}, \gamma_{m-1}^\alpha = \gamma_{m=1}^\alpha (\gamma_1^\alpha, \gamma_2^\alpha) . \tag{6.14}$$

Even though, each node's information density, accumulating its triplet's eigenvalues, grows with increasing the node's number, the ratios of the node's eigenvalues $\gamma_1^\alpha, \gamma_2^\alpha$ for the nodes' internal starting segments of are preserved.

At conditions (6.14) and the limited the eigenvalues ratio, both the space area (6.3c) and its volume (6.11) increase with moving from one triplet to another one along the IN space structure, where the movement takes place along the model's time course $t \to T$. During this move, the node's space is compressed by enfolding its space spot with that of each previous node. However, both the node spot area $f_o$ and the cell spot area $f_o^c$, measured by the same numbers of the enfold bits of the DSS code, are not changed.

It's seen that the total area is determined entirely by a structure of the first triplet with known initial eigenvalue $\alpha_{m=1}$ or time interval $t_{m=1} = \mathbf{a}(\gamma) / \alpha_{m=1}$:

$$F(T_m) = \pi T_m^2 = (\pi t_{m=1})^2 \gamma_{m=1}^{2m} . \tag{6.15}$$

With growing $m \to n/2$, the relative area is growing approaching

$$F(T_m)/F(t_{m=1}) = \gamma_{m=1}^n \quad . \tag{6.16}$$

At forming a triplet, the curvatures changed twice: first, at the eigenvectors' equalization, and second, after their cooperation into a single eigenvector: $|3\alpha_o^i| = |\alpha_o^k|$.

As a result, we come to *three different topological structures* at the cooperation of in curved subspaces with the above eigenvectors (which includes the initial triplet structure with three different eigenvectors being non cooperated yet). Because at the cooperation, the needed transformations arise under the jump-wise control actions, the structures, emerging at their cooperation, have the forms of *discrete* (jump-wise) transitions.

At such jumps, the forming discrete geometrical boundaries between these structures are *topological indicators* of the merged cooperative phenomena. Connection of the cooperative complexity to the curvature implies that the discrete boundaries are *also* the topological indicators of complexity for such a *structure* in its information geometry. (A shared volume of the cooperating structures could be formed by "stitching" of the merged boundaries).

The DSS code-cells, distributed in a fabric of information space geometry (Fig. 2.5), present an elementary form of these discrete structures. Specific code sequence determines the geometrical structure to be built in this information-geometrical space location.

According to the conditions of the cooperation of model's eigenvectors into a triplet (Lerner 2008b, 2010), formation of whole triplet requires the rotation on angle $\varphi_m = \pi$ for each $m$-th triplet and spending the time of rotation $t_m$. Thus, an average angle's speed of rotation for each triplet is $c_m = \pi/t_m = \pi\mathbf{a}(\gamma)\alpha_m\,[rad/\sec]$ at $\alpha_m = [bit/\sec]$.

Since formation of each following triplet also needs rotation on the same angle, its angle's speed would decrease with growing $t_{m+1} > t_m$ and declining $\alpha_{m+1} < \alpha_m$. If both triplets start their formation simultaneously, then the IN's upper level triplet (rotating with maximal angle speed $c_{m=1}$) would finish its formation earlier than the next triplet in this IN's hierarchy would.

In such consideration, the final IN node will rotate with lowest speed $c_{m=n/2} = \pi/T_{m=n/2} = \pi\mathbf{a}(\gamma)\alpha_{m=n/2}$, at $\alpha_{m=n/2} = \alpha_{m=1}/(\gamma_{m=1}^\alpha)^{m=n/2}$. Their speed's ratio $c_{m=n/2}/c_{m=1} = (\gamma_{m=1}^\alpha)^{-n/2}$ decrease with growing $n$ significantly, which at $\gamma_{m=1}^\alpha \cong 3.89, n = 10$ falls to $c_{m=n/2}/c_{m=1} \cong 2.88 \bullet 10^{-4}$.

This means that total IN structure, formed during the relative time $T_{m=n/2}/t_{m=1} = (\gamma_{m=1}^\alpha)^{n/2}$, requires a sequence of rotations of each triplet on angle $\pi$, with sequentially decreasing the angle speeds, having summary angle $\pi n/2$. Such rotating geometrical structure has a spiral form with its invariant cross-section of each triplet's node $f_o = \varepsilon_m^2$ that enfolds the node's cellular DSS code.

Each pair of the triplet's cell-code areas should be turned on angle $\varphi_{2m} = 2\pi$ completing the twist of these two rotations. The node's area $f_o$, in turn, is formed by rotation of the three triplet's double spirals, located on related conic surfaces (Figs.2.4, 2.5), which are joining in the node at the cones' vertexes. All these inner triplet's conic surfaces (with related volumes) are formed during total time $T_m$ and therefore become sequentially enclosed (in this common time) inside the final volume of that conic surface. The node's rotating spiral (carrying only the DSS code) turns on angle $\pi$ between each two subsequent nodes during the time $\Delta t_{12} = t_{m2} - t_{m1}$ and turns on the angle $\pi n/2$ during time $T_{1m} = \sum_{k=2}^{m} \Delta t_{1k} = T_{m=n/2} - t_{m1}$.

The model, having linear speed $c_l [m/\sec]$, determined by ratio of its time and space depended eigenvalues: $c_l = \alpha_i^t / \alpha_i^l = inv$, produces a spiral with the length $L_m = c_l T_{1m}$ during time $T_{1m}$. Such rotating geometrical structure is presumably enclosed within a cylinder having a diameter $l_m = c_l \Delta t_{1m} / 2$ (in a half-length between the nearest nodes), which corresponds to the spiral's rotation on angle $\pi/2$. It's seen that both types of the spiral structures: the rotating spiral-carrier of IN's nodes-code and the spiral, enclosing all inner rotating triplets' $3 \times 2m$ spirals, are produced during the same time $T_{1m}$ needed for creation of total IN geometry.

The $n$-dimensional model, having a manifold of triplet's cell-code areas $m = n/2$, located on the rotating hyperbola, creates an evolving spatial helix dynamic coding structure Fig.2.5.

The numerical examples: for the model with $n = 22, \alpha_{1o} \cong 476.4,$ we get its time $T_{1,22} \cong 1176.85 \sec$ At $t_1 \cong 0.00577 \sec$, and final angle speed $c_{22} = 0.00267 rad/\sec$.

For the model with three triplets and ending control, having $n = 8, \alpha_{1o} \cong 4.36, t_{1o} \cong 0.161$, we get its time and speed: $T_{1,8} \cong 36.7 \sec$, $c_8 = 0.0856 rad/\sec$.

Thus, each cell of the space structure undergoes two kinds of motions:

(1)-with linear space speed in macrodynamic motion $c_l [m/\sec]$ and

(2)-with speed of rotation $c_m [rad/\sec]$ being orthogonal to the macrodynamic motion.

Following these results, the evolution of the surface area, as a function of the time $F = F(T_m)$, is illustrated on Fig.2.10, where each $T_m$ enfolds number of external cellular elements $N_e$ (6.6).

Importance of this function consists on only in revealing the model *space-time' evolution dynamics*, but also in focusing on the observer, possessing such an external space area, where all internal-external interactions take place and the results of Prop.1-5 are employed.

At collective interactions in a limited information space area, the evolution of observer's space area is restricted by maximal available information, which each observer can obtain.

An observer with the time scale (Sec.2.3) enables rotating its structure simultaneously for all nodes.

<u>Footnotes.</u> Turning each eigenvector $\alpha_{i\tau o}$ of triplet's segment on angle $\pm \pi/4$ (by the ending moment $\tau_k^i$) will bring its maximal increment that is measured by the ratio of the eigenvector's modules: $\alpha_{i\tau} / \alpha_{i\tau o} = k_{i\tau}, k_{i\tau} = (\cos(\pi/4))^{-1} \cong 1.154$.

Because all three triplet's eigenvectors undergo sequentially such rotations (needed for their cooperation), the total increment of three rotations is $k_{m\tau} = (k_{i\tau})^3 \cong 1.525$. This number will reduce the initial ratio of the triplet's eigenvalue $\gamma_{mo}$ (prior to the eigenvalue's cooperation) to its potential value after cooperation: $\gamma_m = \gamma_{mo} / k_{m\tau}$. Using an example of the model's computations (Lerner 1999, 2010) with $\gamma_{mo} \cong 3.495$ (which corresponds its minimal value at $\gamma \cong 0.7$) we get $\gamma_m \cong 2.3$. The dynamic invariants $\mathbf{a}(\gamma_{mo}), \mathbf{a}_o(\gamma_{mo})$, defined by $\gamma_{mo}$, are not changed, preserving other the invariants in the rotating macrodynamic process.

## 2.7. Information flow through IN

According to the IN information geometry, each of the IN nodes has the same size of the spot's area and encodes 4 bits-symbols of the IN's information code, while each following node (in the IN hierarchy) enfolds a symbol that carries the code of the previous node's three symbols.

With increasing $m$ nodes-triplets, concentration of the IN initial node's information flow $I_{mo}$ increases in $3^m$ times in each IN's spot area.

That determines information flow going through the IN with $m$ triplets:

$$I_m = I_{mo} 3^m, m = 1, 2, ..., n/2, \qquad (7.1)$$

where $I_{mo} = \alpha_{1o}^{m1} = \partial S_{1o}^{m1}/\partial t$ is initial eigenvalue of the first triplet, which is defined by the related initial entropy derivation.

Growing the concentration (density) of the information flow increases a density of the entropy derivation.

The information flows, moving along the IN nodes, self-join sequentially each flowing triplet with a previous one, which triples the flow density at each IN node. At the IN's final node, such flow density increases in ratio $I_{m=n/2}/I_{mo} = 3^{n/2}$ regarding the initial flow of the first IN triplet.

The IN node's invariants $\mathbf{a}(\gamma_{mo}), \mathbf{a}_o(\gamma_{mo})$ measures the same invariant *quantities* of information for each node, having increasing information *densities* with growing the node number.

Each such a quantity, which accumulates and enfolds a total information from all previous nodes, evaluates the node's information *value* that depends on the node's hierarchical *location* within the IN. Specifically, the node number *m,* identifying its location, indicates the invariant's particular information *value*.

The node invariant quantity is composed from the information contributions of the triplet's three segments and a control, produced by the end of the model's each discrete interval.

The IN's code preserves its invariant four–letters cellular structure, whereas each following node's four letters encode both the previous node code and the two components of its current triplet's node. The cell's sequential number within the triplet coding structure specifies its position and geometry for each triplet, while the *m*-th node's number specifies both the node's location and its geometry within the IN. The IN particular structure carries its total cells' code, as the logical information universal characteristic, which encloses both the generated model's dynamics and geometry and includes all their phenomena.

The information process' flow of information is a carrier of the process' code, and having the code allows decoding this flow-process.

## 2.7a. Information density of the IN code and its evaluation

Let us introduce the notion of an information density $N_b^{sc}$, defined by the number of an information units (bits) that each of this information unit encodes (compresses) from any other source-code( $sc$ ). Since each triplet's bit encodes 3 bits, it information density is $N_b^1 = 3.$ A following triplet also encodes 3 bits, but each of its bit encodes 3 bits of the previous triplet's bits. Thus, the information density of such two triplets is equal to $N_b^2 = 9$, and so on. Hence, for the $m$ -th triplet we have $N_b^m = 3^m$ bits of this $m$ -th triplet which encodes $3^m$ bits from all previous triplet's codes. The IN's final node with $m = n/2$ has $N_b^m = 3^{n/2}$, determined by the process' dimension $n$. The information density, related to the IN's level of its hierarchy, measures also the *value* of information obtained from this level.
For such a code, its information density also measures its value ability.
*For example*, an extension to the architecture of an ARM chip provides the enhanced code density: it stores a subset of 32-bit instructions as compressed 16-bit instructions and decompresses them back to 32 bits upon execution. (See Yiu 2011).

## *2.7. The macrodynamic and cooperative complexities*

Basic complexity measures have been developed in algorithmic computation theory (Kolmogorov1965,1987;Chaitin 1971,1974), important indicators of complexity have been proposed in physics (Bennett 1988,1991); numerous other publications(Solomonoff 1978; Nicolis ,Prigogine 1989; Lopez-Ruiz et all1995; Lopez 1991; Grassberger 1991, Traub et all 1988; Gell-Mann , Lloyd 1996, others) are connected with these basics.
These complexity's measures focus on the evaluating complexity for already formed complex system.
We intend to analyze an *origin* of complexity in an interactive dynamic *process* while its *elements cooperation* into a joint system, accompanied by creation of new phenomena, which in turn, are the *potential* sources of a complexity.
The universality of information language allows a *generalization of* the description of various interactions in terms of the *information* interactions, considered independent of their specific forms and nature. Focus on interactive *informational* dynamics leads us to a study of a *dynamic* complexity resulting from the interactions of information flows, measured by the specific information speeds.
That's why the dynamic information complexity should be connected with the information speeds rather than just with a quantity of information in the above publications.
An intuitive notion of a system's complexity, as a distinction complexity from simplicity, is associated with assembling of the system elements into a joint system during a cooperative process.
This means that the system complexity originates from naturally ability to cooperate and depends on the phenomena and parameters of *cooperative* dynamics.
It has been pointed out repeatedly that algorithmic complexity does not fit the intuitive notion of complexity (Bennett 1988,1991).
The system complexity, emerging from the cooperative dynamics of a *multiple* set of interacting processes (elements), having an adequate information measure, has not been studied yet (in the cited references).
The main questions are: What is a general mechanism of cooperation and the condition of its origin? Does there exist a general measure of a *dynamic* complexity independent of particular physical-chemical nature of the cooperative dynamics with a variety of their phenomena and parameters? How can the dynamic *multi-dimensional* cooperative complexity be defined and measured?
The answers for these questions require a new approach leading us to a *unified notion of dynamic information complexity*, measured in terms of quantities and qualities of information by a corresponding information code.
The *objective* consists of the definition and formulation of the complexity's information measure, the analysis of the complexity's origin in cooperative dynamics, and both analytical and computational measure's connections to the informational dynamic parameters.

Compared to known publications, we analyze the complexity as an *attribute* of the process's *cooperative dynamics,* considering both the phenomenological concept and the formal measure of the complexity.

It is shown that a system complexity depends on both information connections between the interacting elements and the element's number.

Analysis of the *regularities* of collective dynamics, accompanied by a formation of cooperative structures, can be formalized using a variation principle (VP), applied to the informational path *functional* and equations of informational macrodynamics (IMD).

In this Sec. we study both the complexity of information macrodynamic process and the cooperative complexity, arising in interactive dynamics of information flows, which is accompanied with changing of both information flows $\Delta I_{ik}$ (from $I_i$ to $I_k$) and their shared volume $\Delta V_{ik}$ (from $V_i$ to $V_k$):

$$\Delta I_{ik} = I_i - I_k, \Delta V_{ik} = V_i - V_k .$$

The macrodynamic complexity (MC) is defined by an increment of concentration of information in the information flow before and after interaction, measured by the flow's increment per the changed information flow: $MC_{ik} = mes[\Delta I_{ik} / \Delta V_{ik}]$, while the flow's increment is measuring by the increment of entropy speeds: $mes \Delta I_{ik} = \partial \Delta S_{ik} / \partial t$, we get

$$MC_{ik} = (\partial \Delta S_{ik} / \partial t) / \Delta V_{ik} . \tag{7.1}$$

This complexity is determined by an instant entropy's concentration in this volume: $\dfrac{\partial \Delta S_{ik}}{\Delta V_{ik} \partial t}$ (the entropy production), which evaluates the specific information contribution, transferred during the interactive *dynamics* of the information flows.

Complexity (7.1) is measured *after* the considered interaction has held, assuming that both increments of speeds and volumes are known.

To evaluate a complexity arising *during* an interactive dynamics we introduce *an information measure of a differential interactive complexity* $MC_{ik}^{\delta}$, *defined by* the increment of the information flow $-\dfrac{\partial \Delta S_{ik}}{\partial t}$ per a small volume increment $\delta V_{ik}^{\delta}$ (within the shared volume $\Delta V_{ik}$), where

$$MC_{ik}^{\delta} = \frac{\partial H_{ik}}{\partial t} / \frac{\partial \Delta V_{ik}}{\partial t} \tag{7.2}$$

is defined by the ratio of the above speeds, measured by the increments of Hamiltonian and volume accordingly. The $MC_{ik}^{\delta}$ automatically includes both the $MC_{ik}$ and its increment $\delta MC_{ik}$.

The differential complexity (7.2) measures a differential increment of information of interactive elements $i,k$, whose current *information difference* $\Delta S_{ik}$ and shared volume $\Delta V_{ik}$ -*before joining,* would be reduced to increment $\delta S_{ik}$ and volume $\delta V_{ik}^{\delta}$ accordingly after their cooperation during a macrodynamic process.

Applying the IMD, we evaluate both $MC_{ik}$ and $MC_{ik}^{\delta}$ through the VP information invariants, allowing the direct evaluation of these complexity in the bits of information code.

Let us evaluate the complexity of information elements-triplets, currently assembling in a cooperative space distributed hierarchical system (Sec.2.4). Characterizing each triplet's dynamics through their eigenvalues, we assume their inner connection with related geometrical structure.

The differential interactive complexity (7.2) for a triplet, defined at the moment of a three segments eigenvalues' equalization, has a view

$$M^\delta_{i,i+1,i+2} = 3\dot{\alpha}_{i+2,t} / \dot{V}_{i,i+1,i+2}, \qquad (7.3)$$

where $\dot{\alpha}_{i+2,t}|_{t=t_{i+2,t}} = [\alpha^2_{i+2,o} t^2_{i+2,t} \exp(\alpha_{i+2,o} t_{i+2,t})(2 - \exp \alpha_{i+2,o} t_{i+2,t})^{-1} - \alpha^2_{i+2,o} t^2_{i+2,t} \exp 2(\alpha_{i+2,o} t_{i+2,t})] / t^2_{i+2,t}$

$= [\mathbf{a}^2_o \exp \mathbf{a}_o (2 - \exp(\mathbf{a}_o))^{-1} - \mathbf{a}^2_o \exp(2\mathbf{a}_o)] / t^2_{i+2,t} = (\mathbf{a}_o \mathbf{a} - \mathbf{a}^2_o \exp(2\mathbf{a}_o)) / t^2_{i+2,t}, \mathbf{a} = \exp \mathbf{a}_o (2 - \exp(\mathbf{a}_o))^{-1}$ (7.3a)

and

$\dot{V}_{i,i+1,i+2} = \delta V_{i,i+1,i+2} / \delta t, \delta V_{i,i+1,i+2} = V_c \delta t^3, \ \delta V_{i,i+1,i+2} / \delta t = V_c \delta t^2 = V_c t^2_{i+2,t} \delta t^2 / t^2_{i+2,t} = V_c t^2_{i+2,t} \varepsilon(\gamma)^2$ (7.3b).

Here $\varepsilon^2(\gamma) = \varepsilon^2(\gamma^\alpha_1(\gamma), \gamma^\alpha_1(\gamma))$ is an invariant at fixed $\gamma$ and, therefore, fixed $\gamma^\alpha_1(\gamma), \gamma^\alpha_1(\gamma)$; $V_c = 2\pi c^3 / 3(k\pi)^2 tg\psi^o$, where $\psi^o = \pi/6$ is the angle on the vertex of each cone (Fig.2.1), $c$ is a speed of rotation of each cone's spiral, while each of the rotations produces angle $\pi/2$. We get

$$M^\delta_{i,i+1,i+2} = 3(\mathbf{a}_o \mathbf{a} - \mathbf{a}^2_o \exp 2\mathbf{a}_o) / V_c t^4_{i+2,t} \varepsilon(\gamma)^2, \qquad (7.4)$$

which for the joint triplet's segments determines differential complexity $M^\delta_m = M^\delta_{i,i+1,i+2}$.- for any $m$-th triplet-node.

Applying $M^\delta_m$ to a cell with volume $\delta V_{i,i+1,i+2} = \delta V_m$, formed during time interval $\delta t_{i,i+1,i+2} = \delta t_m$, we get $M^\delta_m \delta t_m = 3\dot{\alpha}_m \delta t_m / \delta V_m = 3\Delta \alpha_m / \delta V_m$, where $\Delta \alpha_m$ is an increment of information speed during $\delta t_m$.

The related increment of quantity information at the same $\delta t_m$ is $\Delta \alpha_m \delta t_m = a^\Delta_m = \dot{\alpha}_m \delta t^2_m$, and

$$a^\Delta_m = (\mathbf{a}_o \mathbf{a} - \mathbf{a}^2_o \exp(2\mathbf{a}_o)) \delta t^2_m / t^2_{i+2,t}, \delta t^2_m / t^2_{m,t} = \varepsilon^2_m, t^2_{m,t} = t_{i+2,t}, \qquad (7.5)$$

where $3a^\Delta_m$ measures the quantity of information produced during interaction of three equal eigenvalues within area $\varepsilon^2_m$, and for each invariant $3a^\Delta_m$.

The increment of entropy (in 7.1), determined by the related volume: $M^\delta_m$, is measured by an equivalent quantity information, related to a cell volume $\delta V_m$, during the time $\delta t_m$:

$$M^\delta_m \delta t^2_m = M^\Delta_m = 3(\mathbf{a}_o \mathbf{a} - \mathbf{a}^2_o \exp(2\mathbf{a}_o)) \varepsilon^2_m / \delta V_m. \qquad (7.5a)$$

Information $3a^\Delta_m$ binds these three segments in $\varepsilon^2_m$ prior the action of two impulse controls (Sec.2.6). By the moment of interaction $\tau^{i+2}_k$, three equal eigenvalues have the signs $\alpha_{it}(\tau^{i+2}_k) sign \alpha_{it}(\tau^{i+2}_k) = \alpha_{i+1t}(\tau^{i+2}_k) sign \alpha_{i+1t}(\tau^{i+2}_k) = -\alpha_{i+2t}(\tau^{i+2}_k) sign \alpha_{i+2t}(\tau^{i+2}_k)$.

Since negative eigenvalues $\alpha_{it}(\tau^{i+2}_k) sign \alpha_{it}(\tau^{i+2}_k) = \alpha_{i+1t}(\tau^{i+2}_k) sign \alpha_{i+1t}(\tau^{i+2}_k)$ are stable and the positive eigenvalue $-\alpha_{i+2t}(\tau^{i+2}_k) sign \alpha_{i+2t}(\tau^{i+2}_k)$ is unstable, their interaction leads to instability, associated with a choatic attraction, which is localized within zone $\varepsilon^2_m$. The controls, delivering information $2\mathbf{a}^2_o$, *cooperate* these segments (within $\varepsilon^2_m$) by joining them into a single segment.

Therefore, (7.5a) measures a *cooperative complexity* of this *interactive* three segments, forming *a single node* of $m$-th triplet. The cooperative node is formed by the cell within volume $\delta V_m$, where both the eigenvalues' interaction and cooperation takes place. Since quantity information $2\mathbf{a}_o^2 \cong 1 bit$ of the joint segment from $m$-th triplet's node is transferred to a first segment of following $m+1$-th triplet, the quantity of *binding* information $3a_m^\Delta$ (in (7.5)), being spent on holding $m$-th triplet, is concentrated in the volume $\delta V_m$.

Let us consider $M_{cm}^\Delta = 3a_m^\Delta(\gamma)/[\delta V_m / \varepsilon_m^2]$, evaluating the quantity of information per cell volume $\delta V_m$ related to a cell size $\varepsilon_m^2$. In more simple form, using $M_{cm}^\delta = 3\Delta\alpha_m / \Delta V_m$, $M_{cm}^\delta \delta t_m = 3\Delta\alpha_m \delta t_m / \Delta V_m$, we come to $M_{cm}^\Delta = 3a_m^\Delta / \Delta V_m$, which for each $\Delta V_m$ evaluates $M_{cm}^{\Delta V} = 3a_m^\Delta$.

At $\gamma = 0.5, a_o \cong -0.75, a \cong 0.25$, we get $M_{cmN}^{\Delta V} = 3a_m^\Delta(\gamma = 0.5) \cong -0.897 Nat$ per cell, or $M_{cmb}^{\Delta V}(\gamma = 0.5) \cong -1.29 bit$ per cell-volume that each $m$-th node *conserves* during its formation.

Being produced during the considered interaction (that primarily binds these segments), it measures a *cooperative* effect of the interactions, as the node's *inner cooperative complexity*.

Such a relative cooperative complexity does not depend on the actual cell volume and the number of nodes that the cell enfolds, and the $M_{cm}^{\Delta V}$ invariant quantity is not transferred along the IN nodes' hierarchy. Actually at $\delta V_m / \varepsilon_m^2 = V_c t_m^2$, and a fixed invariant $\varepsilon_m^2$ and volume $V_c$, the increment $\delta V_m$ grows with assembling more nodes.

Since for any cell's volume $M_{cm}^{\Delta V} = inv(\gamma)$ (according to 7.5)), the complexity of each unit of this volume decreases with assembling more cooperating nodes within this volume. This means, with growing the size of a cooperative, the cooperative complexity per its volume decreses in the ratio $M_{m+1}^\Delta / M_m^\Delta = t_m^2 / t_{m+1}^2 = (\gamma_m^\alpha)^{-2}$, while each following $M_{m+1}^\Delta$ enfolds complexity of the previous $M_m^\Delta$.

The absolute value of $\delta t_m = t_m \varepsilon$ grows with increasing $t_{m+1} / t_m = \gamma_2^\alpha$, which leads to $\delta t_{m+1} / \delta t_m = \gamma_2^\alpha$ and $M_m^\Delta = 3a_m^\Delta / \delta t_m \delta V_m, \delta V_m = 3V_c \delta t_m^2, M_m^\Delta = a_m^\Delta / V_c \delta t_m^3 = a_m^\Delta / V_c \varepsilon_m^3 t_m^3$, while $M_m^\delta = a_m^\Delta / V_c \delta t_m^4 = a_m^\Delta / V_c \varepsilon_m^4 t_m^4$.

This confirms the previous relations.

The ratio of the nearest triplet's complexities (7.4) is $M_{m+1}^\delta / M_m^\delta = t_m^4 / t_{m+1}^4$ at

$$(\mathbf{a}_o \mathbf{a} - \mathbf{a}_o^2 \exp(2\mathbf{a}_o)) / V_c \varepsilon(\gamma)^2 = A_M(inv(\gamma)). \tag{7.5b}$$

At $t_m^4 / t_{m+1}^4 = (\alpha_{m+1} / \alpha_m)^4 = (\gamma_{m+1})^{-4}$, satisfaction of (7.5b) . and the IN triplet's node prameter $\gamma_{m+1} = \gamma_2(\gamma) = inv_o(\gamma)$, we get

$$M_{m+1}^\delta / M_m^\delta = \gamma_{m+1}^{-4}, \tag{7.6}$$

which for $\gamma_2(\gamma = 0.5) = 3.89$ takes the values $M_{m+1}^\delta / M_m^\delta \cong 0.00437$. (7.6a)

This means the complexity of $m+1$ node $M_{m+1}^\delta$, while measuring its own complexity, also enfolds and condenses the complexity of a previous node.

<u>Comments 7.1.</u> Let's take into consideration the effect of cooperation, which triples the eigenvalue after cooperation.

By the moment of the triplet's formation $\tau_m$, all its three eigenvalues become equal: $\alpha_{3\tau}^m = \alpha_{2\tau}^m = \alpha_{1\tau}^m$, and at the *moment* of triplet's formation $\tau_m + o$ we come to the following relations for a joint triplet's eigenvalue

$$\alpha_3^m(\tau_m + o) = 3\alpha_{3\tau}^m = \alpha_m,  \tag{7.7}$$

where $\alpha_m$ enfolds all three eigenvalues of $m$-th triplet.

In the IN, the first eigenvalue of $m$-th triplet $\alpha_{1\tau1}^m$ (at the moment $\tau1$, before this triplet is cooperating by the moment $\tau$) becomes *equal* to the last eigenvalue of $(m-1)$-th triplet $\alpha_{m-1}$ (formed after that triplet had cooperated at the moment $\tau_{m-1} + o = \tau1$)); this $\alpha_{1\tau1}^m$ enfolds all three eigenvalues of the previous $(m-1)$-th triplet:

$$\alpha_{1\tau1}^m = \alpha_{m-1};  \tag{7.7a}$$

while for the $m$-th triplet holds true relation

$$\alpha_{3\tau}^m / \alpha_{1\tau1}^m = (\gamma_m^\alpha)^{-1}.  \tag{7.7b}$$

Substituting (7.7b) to (7.7) we have $\alpha_m = 3\alpha_{1\tau1}^m (\gamma_m^\alpha)^{-1}$, and with (7.7a) we get we get ratio

$$\alpha_m / \alpha_{m-1}^m = 3(\gamma_m^\alpha)^{-1}.  \tag{7.7c}$$

The sustained cooperation of the IN eigenvalues requires $\gamma_m^\alpha(\gamma = 0.5) \cong 3.9$, which brings the ratio (7.7c) to the form $\alpha_m / \alpha_{m-1}^m \cong (1.3)^{-1}$. This means, decreasing the triplet's eigenvalues along the IN (according all relations in Secs.2.5) will be satisfied but in three times-less the eigenvalues' ratio, compared to the used: $\alpha_m / \alpha_{m-1}^m = (\gamma_m^\alpha)^{-1}$. Specifically, at $M_{m+1}^\delta / M_m^\delta = (\alpha_{m+1}^4 / \alpha_m^4) / \dot{V}_{m+1} / \dot{V}_m$ and $\dot{V}_{m+1} / \dot{V}_m = \alpha_m^2 / \alpha_{m+1}^2$, $\alpha_{m+1} / \alpha_m = (1/3\gamma_{m+1})^{-1}$, we get

$$M_{m+1}^\delta / M_m^\delta = (\alpha_{m+1}^4 / \alpha_m^4) / \dot{V}_{m+1} / \dot{V}_m = (\alpha_{m+1}^6 / \alpha_m^6) = (1/3\gamma_{m+1})^{-6},  \tag{7.7d}$$

which brings decreasing $M_{m+1}^\delta / M_m^\delta \cong 0.203$ in timeless ratio than in (7.6a). •

Comparing ratio (7.6) with a relative difference of these complexity:
$\Delta M_m^\delta / M_m^\delta = (M_m^\delta - M_{m+1}^\delta) / M_m^\delta = (1 - \gamma_2^4)$, we get $\Delta M_m^\delta / M_m^\delta \cong |0.996|$ at $\gamma_2(\gamma = 0.5) = 3.89$.

It is seent that their difference decreases insignificantly. A relative sum of these complexities:
$\Delta M_{m\Sigma}^\delta / M_m^\delta = (M_m^\delta + M_{m+1}^\delta) / M_m^\delta = (1 + \gamma_2^4), \Delta M_{m\Sigma}^\delta / M_m^\delta \cong 1.0044$ also grows insignificantly.

Considering these complexities within a triplet, we compare the complexities of a possible double cooperation with that of the triple cooperation. We have

$$M_{12}^\delta / M_1^\delta = (\alpha_{12}\delta t_{12} / \delta V_{12}) / (\alpha_1\delta t_1 / \delta V_1) \cong 2(\alpha_2 / \delta V_{12}) / (\alpha_1 / \delta V_1) \text{ at } \delta t_{12} \cong \delta t_1.$$

Since $\alpha_2 / \alpha_1 = (\gamma_2^\alpha)^{-1}, \delta V_{12} / \delta V_1 = (\gamma_2^\alpha)^{-3}$ we get $M_{12}^\delta / M_1^\delta \cong 2(\gamma_2^\alpha)^{-4}$, compared with that for the triplet:

$$M_{123}^\delta / M_1^\delta \cong 3(\gamma_3^\alpha)^{-4}.  \tag{7.8}$$

It is seen that at $\gamma_1^\alpha = 2.215$, $\gamma_2(\gamma = 0.5) = 3.89$, we have $M_{12}^\delta / M_1^\delta \cong 0.083$, and $M_{123}^\delta / M_1^\delta \cong 0.013$.

This means, within a triplet, the complexity decreases more at a double segments'cooperation than that at joining a third segement. However, evaluation of the cooperation between the nearest triplets by the ratio (7.6) (depending on $\gamma_m$) indicates that their cooperative complexity decreases much faster then at the cooperation within a triplet.

The comparison reflects the effect of cooperation, when each following nodes complexity wraps and absorbs the complexity of previous node. Cooperation of each information unit with other binds these units and concerves the bound information. A current decrease of cooperative complexity indicates that more cooperations have just occured.

Cosidering the macrodynamic complexities for each extremal segment:

$M_i^d = \alpha_{it}/\Delta V_{it}, M_{i+1}^d = \alpha_{i+1,t}/\Delta V_{i+1,t}, M_{i+2}^d = \alpha_{i+2,t}/\Delta V_{i+2,t}$, and their ratios

$M_{i+1}/M_i = (\alpha_{i+1,t}/\alpha_{it})/(\Delta V_{i+1,t}/\Delta V_{it})$, $M_{i+2}/M_i = (\alpha_{i+2,t}/\alpha_{it})/(\Delta V_{i+2,t}/\Delta V_{it})$,

where $(\alpha_{i+1,t}/\alpha_{it}) = \gamma_1^{-1}$, $(\alpha_{i+2,t}/\alpha_{it}) = \gamma_2^{-1}$ and $(\Delta V_{i+1,t}/\Delta V_{it}) = (1-\gamma_1^3)$, $(\Delta V_{i+2,t}/\Delta V_{it}) = (1-\gamma_2^3)$,

we get $M_{i+1}^d/M_i^d = \gamma_1^{-1}(1-\gamma_1^3)^{-1}$ and $M_{i+1}^d/M_i^d = \gamma_1^{-1}(1-\gamma_1^3)^{-1}$.

Each of these complexities decreases at the end of the segments' time interval, because of increasing their volumes and decreasing the related eigenvalues, which are ranged according to the VP.

Finally we come to the relation for these $m$-th triple:

$$(M_i^d + M_{i+1}^d + M_{i+2}^d)/M_i^d = \Delta M_{m\Sigma}^d/M_m^d = 1 + \gamma_1^{-1}(1-\gamma_1^3)^{-1} + \gamma_2^{-1}(1-\gamma_2^3)^{-1},$$ (7.8a)

which at $\gamma_2(\gamma=0.5) = 3.89, \gamma_1(\gamma=0.5) = 2.215$ takes value $\Delta M_{m\Sigma}^d/M_m^d \cong 1.05$. (7.8a)

Using invariants relaltions (7.3a) we may express these complexities in the invariant forms:

$$M_i^d = \mathbf{a}/t_i \Delta V_{it}, M_{i+1}^d = \mathbf{a}/t_{i+1}\Delta V_{i+1,t}, M_{i+2}^d = \mathbf{a}/t_{i+2}\Delta V_{i+2,t},$$ (7.9)

Let us compare the summary *macrodynamic* complexities of all triple (7.8a) (related to the complexity of the triplet's fist segment) with the triplet's *cooperative* complexity (7.8) (related to the that for the same first segment).

At $\gamma_2 > 3$, we have

$$1 + \gamma_1^{-1}(1-\gamma_1^3)^{-1} + \gamma_2^{-1}(1-\gamma_2^3)^{-1} \gg 3(\gamma_2)^{-4},$$ (7.9a)

for which, at $\gamma_2(\gamma=0.5) = 3.89, \gamma_1(\gamma=0.5) = 2.215$ we obtain $1.05 > 0.013$.

The results indicate the essential differerence of both types of complexities and the measure of comparative outcome of the segments' cooperation in both a doublet and triplets, while the summary *macrodynamic* complexities measure the complexities prior to these cooperations.

Thus, $M_i^d$ measures complexity of the considered information unit in a *dynamic process*, which produces or consumes the measured information; $M_i^\delta$ measures the unit's information intensity, determined by the quantity of information that this unit intends to spend on the cooperation with other units. When cooperation of this unit with an other occurs, the intensity is deminished, being compensated by that, which binds these units and concerves bound information. A collective unit holds a less information intensity than it was prior to cooperation measured by a summary of each of unit complexities. With more units in the collective, each complexity of an attached unit $M_{i,i+1,i+2}^\delta, i = 1,...,m$ tends to decrease more.

The growing cooperatives intend to spend less information for attracting other units.

However, that is true only for the cooperative, accepting its assembled units under the requirements of a sequential decreasing the information speeds according to MiniMax.

A total (integral) relative *macrodynamics* complexity for entire IN with $m$ triplets is a sum:

$$MC_m^\Sigma \cong m_\gamma , \qquad (7.9b)$$

where for each triplet $m_\gamma$ is approximated by (7.8a). $MC_m^\Sigma$ grows linearly with adding each new triplet.

A total IN's (integral) relative *cooperative* complexity is a sum

$$MC_m^{o\Sigma} \cong \sum_1^m [3(\gamma_2^{-4})]^m = (1-[3(\gamma_2^{-4})]]^m / [1-[3(\gamma_2^{-4})]] , \qquad (7.9c)$$

which is decreasing with adding each new triplet, and at $m \to \infty$, $\gamma_2(\gamma=0.5)=3.89, \gamma_1(\gamma=0.5)=2.215$ holds $MC_m^{o\Sigma} \cong 1.013$. This means, as a total $MC_m^{o\Sigma}$ grows, the complexity of each following cooperation provides a diminishing contribution to complexity of the IN cooperative, and with growing number of such inits, the sum approaches zero. It's seen that the IN's macrodynamics complexity $MC_m^\Sigma$, defined for each non cooperating triplet's segment, in $m$ times higther than the IN's cooperative complexy $MC_m^{o\Sigma}$ as the triplet's number gets bigger. However, at each cooperation, the information quality of a cooperative grows in $3$ times (Sec.2.6).

Formulas (7.9b) and (7.9c) take into account both specific complexities of each triplet and a total number of the IN units.

At unification of triplet's cooperating eigenvectors $\alpha_m = 3\alpha_{i+2}$ in a joint volume $v_m$, each such volume holds this information in the form $\alpha_m v_m$, which we call *information cooperative mass* of this volume, produced at the cooperation

$$M_{vm} = \alpha_m v_m . \qquad (7.10)$$

In more general form of triplet's Hamilotian $\alpha_m = H_m$ , and differential volume $v_m = \delta V_m / \delta t = \dot{V}_m$, we have information mass of the diferential volume

$$M_{vm} = H_m \dot{V}_m . \qquad (7.10a)$$

Using connection of entropy derivation $\partial \Delta S_m / \partial t = -H_m$ with related entropy's divergence $div \Delta S_m$ for the same volume $v_m$, in the form

$$\partial \Delta S_m / \partial t = c_m div \Delta S_m , \qquad (7.11)$$

where $c_m$ is a liner speed of the $div \Delta S_m$ at the cooperation of $m$-the triplet, we can determine the information mass through this divergence:

$$M_{vm} = -(c_m div \Delta S_m) \dot{V}_m . \qquad (7.11a)$$

Considering the ratio of the information mass for a nearest triplets

$$M_{vm} / M_{vm+1} = \alpha_m / \alpha_{m+1} (v_m / v_{m+1}), \qquad (7.11b)$$

where $\alpha_m / \alpha_{m+1} = 3(\gamma_{m+1}^\alpha)^{-1}$, and according to (7.3b): $v_m / v_{m+1} = t_m^2 / t_{m+1}^2$ at $\varepsilon(\gamma)^2 = inv$, we get $v_m / v_{m+1} = \alpha_{m+1}^2 / \alpha_m^2 = (1/3\gamma_{m+1}^\alpha)^2$ and

$$M_{vm} / M_{vm+1} = 1/3\gamma_{m+1}^\alpha \qquad (7.11c)$$

which at $\gamma_{m+1}^{\alpha} \cong 3.9$, leads to growing the mass in 1.3 times with adding each following IN's triplet.
The thriplet cooperative complexity for the same volume is
$$M_m^{\delta} = \dot{H}_m / \dot{V}_m. \tag{7.12}$$
Information mass (7.10), related to complexity (7.12), is connected to ratio of the Hamiltonians:
$$M_{vm} / M_m^{\delta} = H_m / \dot{H}_m, \tag{7.12a}$$
which for $H_m = \alpha_m, \dot{H}_m = \dot{\alpha}_m$, brings
$$M_m^{\delta} / M_{vm} = \alpha_m / \dot{\alpha}_m. \tag{7.12b}$$
The last relation is connected to curvature $K_{\alpha}^m = -3\alpha_m / \dot{\alpha}_m$ for the triplet in the form
$$K_{\alpha}^m = -3 M_{vm} / M_m^{\delta}. \tag{7.13}$$
Curvarure $K_{\alpha}^m$ decsribes a curving phase space at locality of the widows within volume $v_m$, forming at the cooperation of three triplet's eigenvectors. This curvature follows from classical Gaussian curvature in a Riemann space (Einstein 1921), which is defined via a fundamental metric's tensor $\sqrt{g}$, describing a closeness of the vectors in this space, in the form
$$K_m^{\alpha} = (\sqrt{g})^{-1} \partial(\sqrt{g}) / \partial t. \tag{7.14}$$
For the considered eigenvectors in the information phase space, metrical tensor $\sqrt{g}$ is expressed (Lerner 2007, 2010a) via the matrix's components of three eigenvectors before and after coperation. The information, curried by the eigenvectors and localized in a space, generates an increment of tensor $\sqrt{g}$, which allows us measuring an information, produced at an interaction of the eigenvectors. Specifically, it has shown that at the triple cooperation, the model's tensor acquires the form $\sqrt{g} = (\alpha_m)^{-3}$, where $\alpha_m$ is eigenvalue of the cooperating triple.

This determines $K_{\alpha}^m = -3\dot{\alpha}_m \alpha_m^{-1}$, which according to (7.13), is a result of both the cooperation and the memorized information mass, or a cooperated information mass, which generates complexity.
The cooperation decreases uncertainty and increases information mass at forming each triplet according to (7.11c). Thus, at the cooperation, accompanied by the decreases of triplet's eigenvalues, complexity (7.7d) declines in much higther ratio than information mass (7.11c) increases, leading to lovering the curvalure of the cooperated IN's structure. The negative curvature (7.13) characterizes a topology of the space area where the cooperation takes place.
The information mass, defined by the cooperating eigenvalue (7.10), can be encoded by the IN triplet code for each its volume (Secs.2.4, 2.5), as well as the cooperative complexity (7.4).
Therefore, curvature in (7.13) can also be encoded using both the complexity's and mass' codes.
The code's density (Sec.2.6a) is growing at each triple cooperation for each encoded information mass. This means that for increasing mass (7.11c), the code's concentration in the mass raises, but lesser than the growing code's density. According to (Einstein 1921), multiplication of a mass on $\sqrt{g}$ determines a mass density. In our case, where this tensor is expressed via the model's eigenvector at the cooperation ($\sqrt{g} = (\alpha_m)^{-3}$), its multplication on information mass (7.10), leads to the mass *density* $M^*_{vm}$, which acquires the form
$$M^*_{vm} = (\alpha_m)^{-3} \alpha_m v_m = (\alpha_m)^{-2} v_m. \tag{7.15}$$

In the simulated IN hierarchy (Lerner 1997, 2001)(see also Fig.2.4), the values of cooperating eigenvalues $\alpha_m$ decrease with a growing number of triplets $m \to n/2$, (following also from Sec.2.5), which leads to an increasing of $M*_{vm}(m)$. *Finally, the information mass, following from (7.13), emerges as a curved information space per its coopertaive information complexity.*

Using relation (7.11) we will evaluate a maximal speed $c_{mo}$ for an elementary single cooperation. Applying invariant $a_o$ for evaluation $\partial \Delta S_m / \partial t = |a_o|/t_m$ we have $c_{mo} = t_{mo} div \Delta S_m / |a_o|$, where we estimate $t_{mo}$ by minimum admissible time interval $t_{mo} \cong 1.33 \times 10^{-15}$ sec (defined by the minimal time-interval of the light wavelenght $l_{mo} = 4 \times 10^{-7} m$), and estimate the normalized divirgence by realtion by a structural invariant of minimal uncearainty 1/137 (Lerner 2010a):

$$div * \Delta S_m = div \Delta S_m / |a_o| \approx 1/137 \tag{7.16}$$

we get the maximal information speed $c_{mo} \approx 1.03 \times 10^{17} Nat / \sec$. (7.17)

This maximum restricts the cooperative speed and a minimal information curvature at other equal conditions. From (7.16) it also follows that a *bound into space* information ($div * S_i$) (by an elementary cooperation) limits the maximal speed of incoming information, imposing an *information* connection on the time and space. Unbound information (a code's symbol) would not have such limtation. The ratio of speed (7.17) to the speed of ligth $c_o$:

$c_{mo}/c_o \approx 0.343 \times 10^9 Nat/m = 0.343 gigaNat/m$ (in a light's wavelength meter) limits a maximal information space speed. In this case, each light wavelength carries $\cong 137$ Nats during $1.33 \times 10^{-15}$ sec, which are delivered with speed of light.

The physical mass-energy that satisfies the law of preservation energy (following the known Einstein equation), is distinguished from the information mass (7.10), which does not obey this law.

*Connection to Kolmogorov Complexity*

Algorithmic Kolmogorov (K) complexity (Kolmogorov 1965) is measured by the relative entropy of one object (*k*) with respect to other object (*i*), which is represented by a shortest program in bits.

The $MC_{ik}^{\delta}$ complexity measures the *specific quantity of information* (transmitted by the relative information flow), required to join the object *i* with the object *k*, which can be expressed by the algorithm of a minimal program, encoded in the $MC_m^{\delta}$ (IN) communication code. This program also measures a "difficulty" of obtaining information by *j* from *k* in the transition *dynamics*.

$MC_{ik}^{\delta}$ represents the information measure between *order and disorder* in stochastic dynamics and it can detect determinism amongst the randomness and singularities. Because the IPF has a limited time length and the IPF strings are *finite*, being an upper bound, the considered cooperative complexity is *computable* in opposition to the *incomputability* of Kolmogorov's complexity.

*The MC-complexity is able to implement the introduced notion and measure of information independent on the probability measure by applying the IN information code for the object's processes.* In the IPF-IMD, an object is represented by random processes, while their observations are measured by dynamic processes; and this approach's aim is to reveal the object's information in a form of its *genetic code. This approach* differs from both the Shannon information of an object's *random events' observation* and the Kolmogorov encoding of *an individual* object's description (in a binary string) by a shortest algorithm; the algorithmic complexity is not required the description of probability function. This is also not needed for the cooperative complexity.

## 2.8. Restoration of the optimal process' parameters

The process basic parameters $(n,\gamma)$ we restore using the observing process's limited variables:

(1)-average maximal frequency $f_o$, identified through its maximal $f_{omx}$ and minimal $f_{omn}$ values, which are connected with their time intervals: $f_{omx} = t_{omx}^{-1}, f_{omn} = t_{omn}^{-1}$ ;

(2)-and a total time $T_p$ of the process' existence.

Applying theorem of evaluation of the real and imaginary parts of eigenvalues $\lambda_o$ (for a complex linear opertator, Korn 1961), we come to relation
$\operatorname{Re}\lambda_o \cong 1/2(f_{omx}^{-1} + f_{omn}^{-1}), \operatorname{Im}\lambda_o \cong 1/2(f_{omx}^{-1} - f_{omn}^{-1})$, from which we get
$\operatorname{Im}\lambda_o / \operatorname{Re}\lambda_o = \gamma_o = (f_{omx}^{-1} - f_{omn}^{-1})/(f_{omx}^{-1} + f_{omn}^{-1})$
or using $f_{omx} = t_{omx}^{-1}, f_{omn} = t_{omn}^{-1}$ , we come to $\gamma_o = (t_{omx} - t_{omn})/(t_{omx} + t_{omn})$, being evaluated by these maximal and minimal time intervals.

This $\gamma_o = \gamma$ will be constant for the identified process (Sec.2.3), for which the partucular VP is applied. That $\gamma$ allows finding $\gamma_2^\alpha(\gamma)$ from Sec.2.4 and then getting the information invariants from (1.8.4.1.8.4a). Formulas $T_p / t_o = [\gamma_2^\alpha(\gamma)]^{m=n/2}$ and $t_o = 1/2(t_{omx} + t_{omn})$ at the known $\gamma_2^\alpha(\gamma)$ and $f_{omx} = t_{omx}^{-1}, f_{omn} = t_{omn}^{-1}$ allows finding the process' dimension $n$.

*Example.* Let us have $T_p \cong 150 \sec, f_{omx} \cong 23.25(t_{omx} \cong 0.043), f_{omn} = 21.3(t_{omn} \cong 0.047)$.

We get $t_o \cong 0.045$ , $\gamma = 0.044/0.09 = 0.488 \cong 0.5$ , from which it follows $\gamma_2^\alpha(\gamma) = 3.8955$.

Using $T_p / t_o = 150/0.045 = (3.8955)^{n/2}$ we come to $n \cong 12$.

From that, we get the information density of the proccess code $N_b^m = 3^6 = 729$, which means that each bit of the IN's final code encodes 729 bits .

If one ecodes some initial code with its symbols (bits) into four symbols (bits) of the IN's first node, having an initial information density $N_b^r$, then through building the optimal IN of dimension $m$, one can increase the information density of its code up to $N_b^r N_b^m = N_b^r 3^m$ times. This method opens practically unlimited compression via a huge number of the IN dimension, restricted only by a maximal admissible IN's dimension, sustaining its formation: $n^m \cong 300$ ( Lerner, 2001), or $N_b^m = 3^{150}$.

To apply this method, an external code should be encoded (or compressed) into a first (starting) IN node's code, as an input of a computer program, which for given $(n,\gamma)$ produces the IN final code that encodes the external code with the requested information density.

The computer operations (Lerner, 2010, 2013), at a known $(n,\gamma)$, allow us to find the optimal sequence of the example's ranged $(\alpha_{io}, \alpha_{it}), i = 1,...,12$, the discrete intervals $(t_{io}, t_{it})$, and to resore the optimal process for each extremal segments $x_{it}(x_{io}, t)$, with a sequence of the applied optimal control for both $v(t_i), u(t_i)$, including their stepwise and impulse components.

These computations also restore the IN structure with its node's hierarchy, including space-time geometry of its external surface, curvature, the macromodel's dynamic and geometrical border, and the coooperative complexities.

The methodology is applicable to any system.

Assuming the first initial condtions $x_{1o}=1$, the following $x_{2o}, x_{3o}$ $x_{2o}, x_{3o}$ will be found from relations Sec.1. 6 (1.6.6b,e) others, applying their ranged invariant sequence.

The covariation matrix on each time interval $t_k: r_{it}(t_k) = t_o \exp(-2\alpha_{io}(t_k)t_k) = t_o \exp 2\mathbf{a}_o(\gamma)$ will be found at the initial coefficient of covariation $r_o(t_o) = E[x_o^2(t_o)] \cong t_o$.

This means, it's not necesserily to measure the initial entropy functional and all parameters that determines the considered stochastics and informational macrodynamics.

Increments of information at fixed time interval $\Delta t_*$, taken on a first (starting) level of the IN: $\Delta S_1(\Delta t_*) = \alpha_{1t^*}\Delta t_*$ and on its $m = n/2$ level: $\Delta S_m(\Delta t_*) = \alpha_{mt^*}\Delta t_*$ at the same time interval (according to time scale, Sec.2.4), differ in $\Delta S_1(\Delta t_*)/\Delta S_m(\Delta t_*) = \alpha_{1t^*}/\alpha_{mt^*} = (\gamma_2^\alpha)^m$ times. This means, the IN's ogranized structure essentially minimizes its initial information in the ratio $(\gamma_2^\alpha)^m$. However, each bit of this minimum is more informative (in the terms of ratio $(\gamma_2^\alpha)^m$), containing a highest density of its information code. From these, it follows that the VP basic principle: "get maximum information from its minimum" brings more information than applying a principle of maximizing an absolute information. Because, selecting a minimum among available maximums (or a maximum among available minimums) means getting such a minimum from a *more organized information system*, which provides a valuable information with a highest code's density.

## *2.9. The IN's interactive information structure as an object-observer*

The self-organized dynamic information structure represents an information extractor (an object), considered as an information observer (Lerner, 2011, 2012), which enables self-generate this structure during the EF-IPF functional transformations of the environmental information, forming the object's dynamically changed *geometrical* boundary (Sec.2.3). (A primary *information boundary* is created by a separation of the EF *uncertainty* of a random process from a *certainty* of the IPF information dynamic information process). The observed information might include random information *obtainable from other observers* that presumes the observers' interaction, which for each observer could be located on its geometrical boundaries-windows.

Benatti et all 2003, have shown that even presence a common environmental noise can create an indirect interaction through the subsystems (observers) entanglement (Sec.1.11).

Information spectrum on the observer's border models a wide spectrum of potential physical frequencies, whose vibrating strings reproduce a huge collection of different physical objects, from varieties of physical atomic, subatomic particles, molecular, macromolecular, to a wide range of different macrosystems.

Each IN surface is composed by the cells each of them holds information unit of the DSS code with the invariant measure $\mathbf{a}_o \cong 1 bit$ (at $\gamma \to 0$). Each IN node on the surface encloses four cells $m_c = 4$ (each with $f_o^c = 1 bit$ of the DSS code), and total surface area $F$ contains information $F_{im} = m_c f_o^c S_m$, where $S_m$ is information delivered by $m$ numbers of the IN nodes on this surface. Assuming that each observer's interacts with environment via the node's surface area, we get all observers' information available for the interaction $S_m = 1/4 F_{im}$, with total cells' number on surface $N = F_{im}/f_o^c$. Here $S_m$ depends only on the cell's surface area and is not dependable on the surface curvature (Sec. 2.6a) (because each cell holds 1-bit information independently of the cell's area curvature).

We also suppose that other observers also interact with the considered observer through *each* of these observers' single IN node. Then, we can evaluate the number of the interacting observers by $N_{em} = 1/4 N$ with total maximal information collected during the interactions $S_{em} = N_{em} = 1/4 N f_o^c$ bits.

Since each node encloses a control information, in addition to the triplet's information in three bits, which is enfolded in the IN cooperative binding, only this additional information can be transferred to the interacting observers at each interaction. This means, the observer's interaction can be considered as their mutual control, which confirms our previous results (Lerner, 1973, 2008a).

Here we assume that each interacting observer acts according to the VP minimax principle, keeping a balance of the consumed (external) and internal information (uncertainty-certainty). Specifically, during each extremal movement, until entrance on a random widow (at a quantum information locality), an observer is preparing itself for the acquisition of new information. Getting ready at the moment $t_k$, indicating by a jump of information speeds (Sec.1.5,1.11), the observer generates a step-down control in the next moment $t_k + o$ which initiates the extraction a maximum information with aid of impulse controls on each following window, which is predicted by the VP.

Therefore, the observer *a priory* decides when he/she is ready for acceptance new information and what quantity and quality of this information is needed.

This indicates the observer's *predictable actions* and intentions for future information acquisitions.

However, a quantum uncertainty at a locality of this actions (before entanglement takes place-at reaching the dynamic constraint, Sec.1.11), makes observer's intention simultaneously both definitive and not, which leads to a possibility of both getting observation (through extraction of information) and not doing it. (Perhaps, reflecting observer's free will).

Classical intention for observation defines the moment $t_k^i$ of turning constraint on, after which observation should start. The generated control overcomes this uncertainty. The observer's dynamic model, which does both conversion of an observed random process in the related macrodynamics, cooperation of the multiple macrodynamics in the IN's hierarchy, and generation of a generic code of the observed random process, works as an observer's *operating system*.

The observers' interactions "inherit and originate evolution of developmental innovation, being a path for evolution of novel adaptations in complex multi-cellular organisms" (Bodyaev 2011). (See Evolution Dynamics in Part3).

The macromodels inherit these and other peculiarities above through acquisition of a maximum needed information along the IN hierarchy, which is associated with increasing of valuable information and emergence of cooperative complexity (Lerner 2011).

In the environment with limited information sources, the observers compete for maximum available information, which leads to selecting the ones with more valuable information, having a maximal dimension and cooperative complexity.

*2.10. The causal-consecutive relationships in the IN dynamics and geometry*

The IMD embraces both the individual and collective regularities of the information model and its elements, forming multiple connections and variety of IN information cooperative networks, with the growing concentration of information and its volume.

The interactive dynamics of the IN nodes are able to produce new information.

The IN, identified on a particular process, characterizes its dynamic hierarchical information structure of collective states, representing a collective dynamic motion.

Both cooperation and collectivization are driven by the intention to get maximum information from its available minimum, which is expressed by the VP (via its constrain, directed on joining of dynamic motions on the extremal segments).

The impulse control, implementing the cooperation, switches the macroprocess from the extremal segment, satisfying a current entropy minima to an extremal, satisfying its decrees (after an influx of a local maxima between the extremals brought by each impulse control), and spends it on the cooperation. Local entropies are enclosed through sequential cooperation of the IN nodes, creating an information structure, which condenses the total minimal entropy being produced (evaluated by the VP functional's derivation (Sec.1.5) at the end of each segment). The VP coordinates a balance between the structural information of the IN and the amount of external information.

An influx of the VP entropy maxima coincides with the model's maxima of optimal acquisition of external information used for an external observation and prediction of the process' observed phenomena. The increasing structural information reduces the amount of external information needed. New information arises at the transference from the current segment, where existing information is accumulated, to a newly formed segment. The transformation is implemented by the control action, which, being initiated by the accumulated information, interacts with the microlevel information and transfers the emerged (or renovated) information to a new segment. Such new information is a result of an interaction between existing (accumulated) information and the information currently delivered through the microlevel's windows.

The final IN node collects a total amount of information coming from all previous nodes.

The dynamic process within each segment's extremal, where the VP applies, is reversible. Irreversibility arises at the random windows between segments, before the segments are joined by the controls at the DP.

The triple formations are analogous to three-critical phase transformations of the second order with a specific connection of the kinetics, diffusion and the symmetry order (Lifshitz, Pitaevsky 1979; Stainley 1980). The consolidated information states are asymmetrical; they cannot coincide using the symmetrical transformations (Sec.2.6). The impulse controls consolidate the asymmetrical local instable irreversible macrostates into stable cooperative structures at each new discrete interval. The controllable self-cooperation generate the IN dissipative structures (Nikolis, Prigogine 1977, Lerner 2006b).

Assembling of each of the three segments into a triplet is accompanied by a local instability and arising of chaotic oscillations, which potentially initiate a chaotic resonance.

The connected triplet's chain generates a *collective* resonance, where the contributions of all triplets sound in unison. This procedure synthesizes a *harmony* of the assembled triplets (Lerner 2006).

The IMD hybrid processes include microlevel irreversible stochastics, reversible macrolevel dynamics, quantum dynamics between them, and cooperative dynamics.

The VP allows establishing not only information connections between the changes, identified by specific sources and related information measures (Sec.1.11), but also *finding the regularities* of these connections in the form of a complex *causal-consecutive relationship* and an information network, carried by EF-IPF analytics and the following logic operations and implemented through building IN and DSS.

Specifically, there are three forms of causalities within the IN (Fig.2.1, 2.3-2.6):

(1) Local –within each VP extremal segment when an information force (and a control at the segment's beginning) initiates (causes) a related information flow along the segment. This is a symmetric deterministic causality, accompanied by a potential reversibility in the time of this local causal action;

(2) Interim –within a "window", initiated by some external interactive actions (including a control), which cause the effect at next extremal segment ad joint to the window. This is an asymmetric casualty, accompanied by irreversible time course and generally nondeterministic causal relationships;

(3) Global–arises along the IN at the information transformation from each previous to the following node. This causality includes both the local causality (within each path between the nodes) and the interim causality (at each transformation from the node to the above path). However, the main specific of this causality consists of arising *total* sequential causal relationships along the IN hierarchy: from the IN starting events to first triplet's node, which joins the first three causes, and then to the following nodes, which sequentially bind all previous node information, and, finally, encloses it in the IN ending node, which accumulates all IN causalities.

Thus, a mutual connection of symbols in the information process leads to a self-collectivization them into the segments and then to a *self-organization* in the IN hierarchy, with its nodes, as an information objects, having the causal-consequence relationships.

**References to Part 2**

Badyaev A.V.Origin of the fittest: link between emergent variation and biology evolutionary change as a critical question in evolutionary, *Proc. R. Soc. B* 278, 1921-1929 , 2011.

Bennett C. H. *Logical depth and physical complexity*, in: *The Universal Turing Machine*, R. Herken (Ed.), 227-258, Oxford University Press, 1988.

Bennett C. H. *How to define complexity in physics, and why*, in: *Complexity, Entropy and Physics of Information,* W. H. Zurek (Ed.),137-148, Addison Wesley, New York, 1991.

Chaitin G. J. Computational Complexity and Godel's incompleteness theorem, *SIGACT News*, 9:11-12, 1971.

Chaitin G. J. Information-theoretic computational complexity, *IEEE Trans. Information Theory*, IT-20:10-15, 1974.

Gell-Mann M. Lloyd S. Information Measures, Effective Complexity and Total Information, *Journal Complexity*, 2:44-52, 1996.

Grassberger P. *Information and Complexity Measures in Dynamical Systems*, in: *Information Dynamics*, Atmanspacher H, Scheingraber H. (Eds.), 15-33, Plenum Press, New York, 1991.

Einstein A. *The meaning of Relativity*, Princeton University Press, Princeton, 1921.

Kolmogorov A. N. Three approaches to the quantitative definition of information, *Problems Information. Transmission*, 1 (1): 1-7, 1965.

Kolmogorov A. N. *Information Theory and Theory of Algorithms*, Selected Works, Nauka, Moscow, 1987.

Koch K., McLean J., Berry M., Sterling P., Balasubramanian V. and Freed **M.A.** Efficiency of Information Transmission by Retinal Ganglion Cells, *Current Biology*,16(14): 1428-1434, 2006.

Korn G.A. Mathematical Handbook for Scientists and Engineers, MGraw-Hill, New York, 1961.

Landau L.D and Lifshitz E.M. *Mechanics,* Nauka, Moskow, 1965.

Lerner V.S. *Superimposing processes in control problems*, Stiinza, Moldova Acad., 1973.

Lerner V. S. Macrosystemic Modelling and Simulation, *Journal Systems Analysis-Modeling-Simulation*, 28 (5):149-184, 1997.

Lerner V. S. Information Geometry and Encoding the Biosystems Organization, *Journal of Biological Systems*, 13 (2):1-41, 2005.

Lerner V.S. Macrodynamic cooperative complexity in Biosystems, *Journal Biological Systems*, **14**(1):131-168, 2006a.

Lerner V.S. About the Biophysical Conditions Initiating Cooperative Complexity, Letter to Editor, *Journal of Biological Systems*, 14(2), 315-322, 2006b.

Lerner V. S. *Information Systems Analysis and Modeling:An Informational Macrodynamics Approach*, Kluwer, 1999.

Lerner V. S. Building the PDE Macromodel of the Evolutionary Cooperative Dynamics by Solving the Variation Problem for an Informational Path Functional, *International Journal of Evolution Equations* 3(3):299-355, 2009.

Lerner V.S. *Information Path Functional and Informational Macrodynamics*, Nova Sc.Publ., New York, 2010a.

Lerner V.S. The entropy functional's and the information path functional's basics with the thermodynamics and cooperative information dynamics applications, *Journal Integration: Mathematical Theory and Applications*, 2(1):1-38, 2010b.

Lerner V.S. The mathematical law of evolutionary informational dynamics and an observer's evolution regularities, *arXiv*:1109.3125,09-14-2011.

Lerner V.S. An observer's information dynamics: Acquisition of information and the origin of the cognitive dynamics, *Journal Information Sciences*, 184, 111-139, 2012.

Lerner V.S. Hidden Information and Regularities of an Information Observer: A review of the main results R, *arXiv*:1303.0777, 2013.

Lifshitz E. M., Pitaevsky L. P. *Physical Kinetics,* Nauka, Moskow, 1979.

Lopez L. R., Caufield L. J. A Principle of Minimum Complexity in Evolution, *Journal Lecture Notes in Computer Science*, 496: 405-409, 1991.

Nikolis G., Prigogine I. *Self-organization in Nonequilibrium Systems*, Wiley, New York, 1977.

Nicolis G. and Prigogine I. *Exploring complexity*. W. H. Freeman, New York, 1989.

Reuveni I., Saar D., Barkai E. A Novel Whole-Cell Mechanism for Long-Term Memory Enhancement, *PLoS ONE,* 8(7), e68131. doi:10.1371/journal.pone.0068131,2013.

Shannon C.,E., Weaver W. *The Mathematical Theory of Communication*, Illinois Press, Urbana, 1949.

Schrödinger E. *The present situation in quantum mechanics*, Naturwissenschaften,1935.

Solomonoff R. J. Complexity-based induction systems: comparisons and convergence theorems, *IEEE Transactions on Information Theory*, 24: 422-432, 1978.

Stainley H. E. *Introduction to phase transformation and critical phenomena*, Oxford Univ. Press, Oxford, 1980.
Stratonovich, R.L. *Theory of Information*, Soviet Radio, Moscow, 1975.
Traub J. F., Wasilkowski G. W, Wozniakowski H. *Information-Based Complexity,* Academic Press, London, 1988.
Yiu J. *The Definitive Guide to the ARM Cortex-M0*; 2nd Edition, Newnes, 2011.
**Figures**

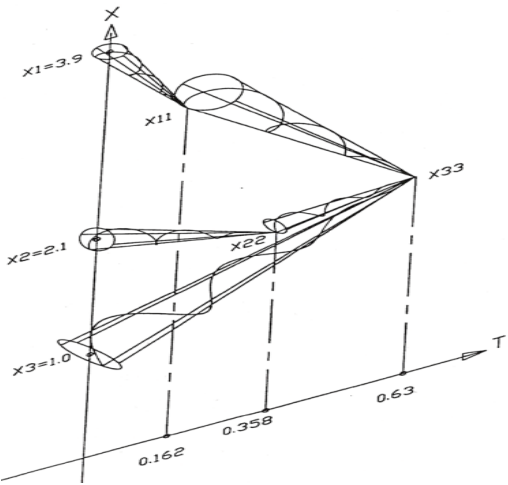
Fig. 2.1. Forming a triplet's space structure.

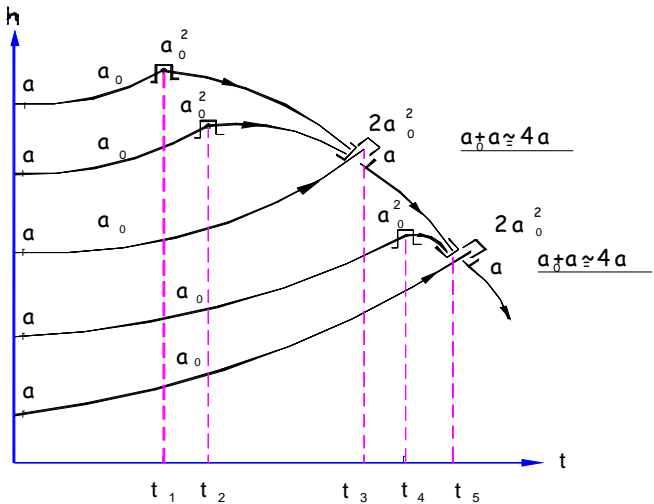
Fig.2.2. The information structure of cooperating triplets' segments with applying impulse controls.

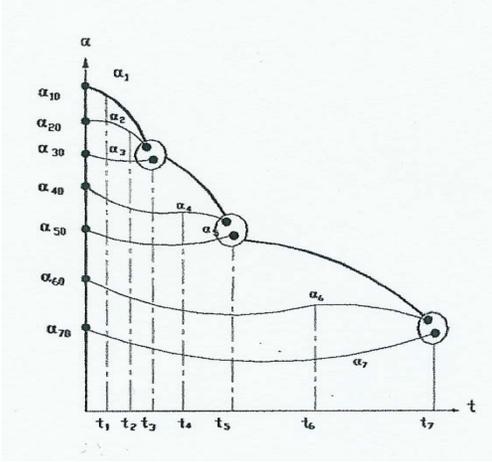

Fig. 2.3. The equalization of the model's eigenvalues for the corresponding eigenvectors during the optimal movement with the triplet node's formation at the localities of the triple cones vertexes' intersections; $\{\alpha_{io}\}$ is a ranged string of the initial eigenvalues, which are cooperating (during the time dynamics) into the triplets, formed around the $(t_1, t_2, t_3)$ locations.

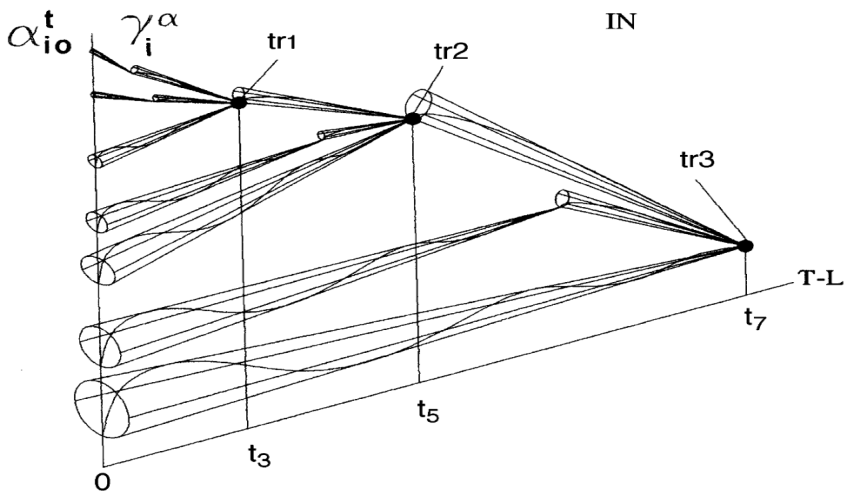

Fig. 2.4. The IN time-space information structure, represented by the hierarchy of the IN cones' spiral space-time dynamics with the triplet node's (tr1, tr2, tr3, ..), formed at the localities of the triple cones vertexes' intersections, where $\{\alpha_{io}^t\}$ is a ranged string of the initial eigenvalues, cooperating around the $(t_1, t_2, t_3)$ locations; T-L is a time-space.

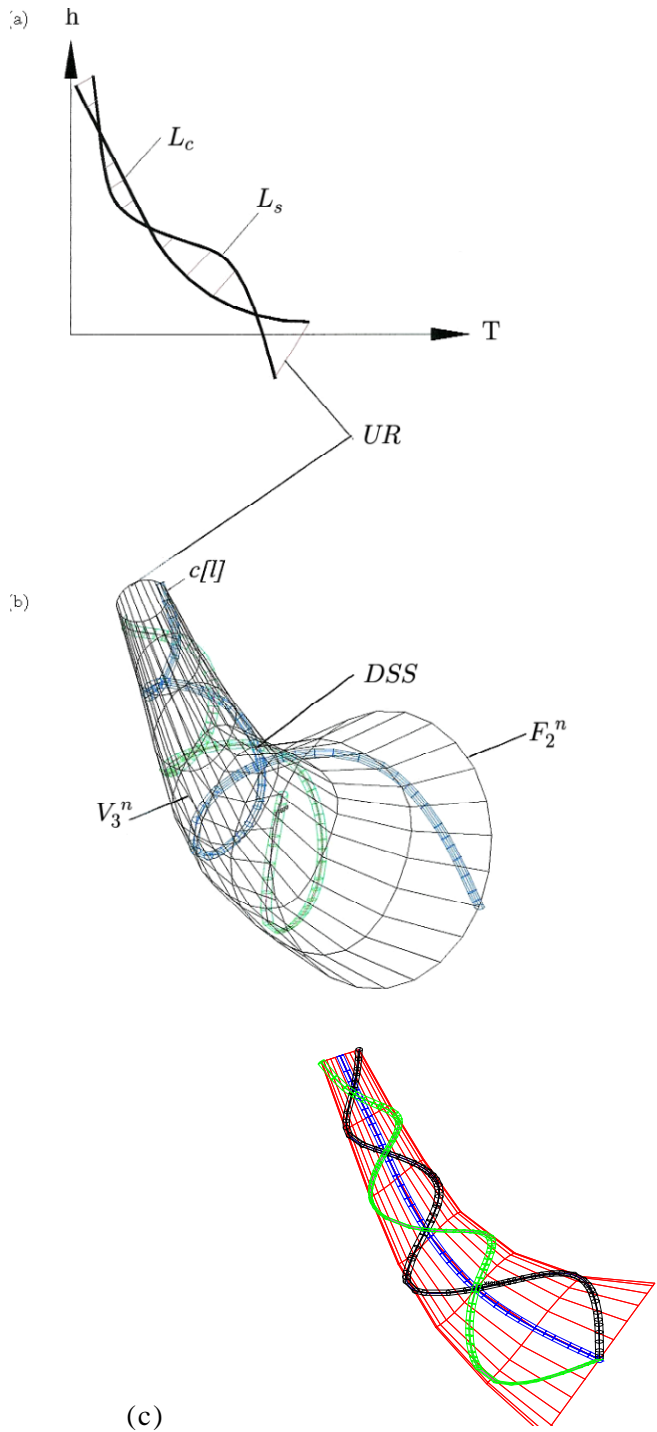

Fig 2.5. (a) Simulation of the double spiral cone's structure (*DSS*) with the cell (c[l]), arising along the switching control line *Lc*; with a surface $F_2^n$ of uncertainty zone (*UR*) (b), surrounding the *Lc*-hyperbola in the form of the *Ls*-line, which in the space geometry enfolds a volume $V_3^n$ (b,c).

(c) Simulation of the double spiral code's structure (*DSS*), generated by the IN's nodes: the central line models the IN node's information cells; the left and right spirals encode the IN's states, chosen at the DPs by the IC control's double actions.

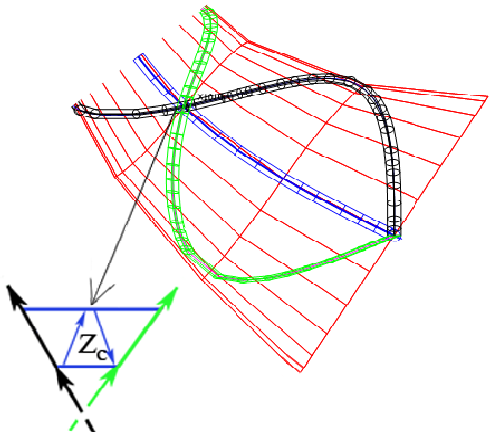

Fig.2.5d. Zone of cells $Z_c$ formed on the intersections of opposite directional spirals which produces each triplet's DSS code

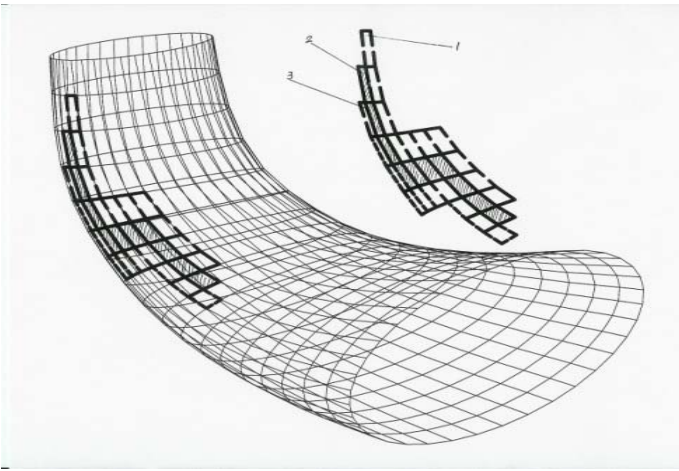

Fig. 2. 5. Structure of the cellular geometry, formed by the cells of the DSS triplet's code, with a portion of the surface cells (1-2-3), illustrating the space formation.

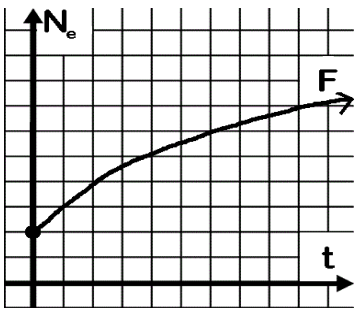

Fig.2.6. The number of external elements $N_e$ as a function of observer's external surface $F$.